\documentclass{article}

    \PassOptionsToPackage{numbers, compress}{natbib}
 \usepackage[preprint]{neurips_2026}

\usepackage[utf8]{inputenc} 
\usepackage[T1]{fontenc}    
\usepackage{hyperref}       
\usepackage{url}            
\usepackage{booktabs}       
\usepackage{amsfonts}       
\usepackage{nicefrac}       
\usepackage{microtype}      
\usepackage{xcolor}         

\usepackage{hyperref}       
\usepackage{bibentry}
\usepackage{amsfonts}
\usepackage{multirow}
\usepackage{amsmath}

\usepackage{graphicx}
\usepackage{epsfig}
\usepackage{color}
\usepackage{epstopdf}
\usepackage{rotating}
\usepackage{caption}
\usepackage{float}
\usepackage{multicol}
\usepackage{amssymb}
\usepackage{graphicx}
\usepackage{bbding}
\usepackage[linesnumbered,ruled,vlined]{algorithm2e}
\usepackage{balance}
\usepackage{microtype}
\usepackage{booktabs}
\usepackage{multirow}
\usepackage{amsmath}
\usepackage{mathtools}
\usepackage{etoolbox}
\usepackage{cases}
\usepackage{enumitem}
\usepackage{xcolor}
\usepackage{cases}
\usepackage{dsfont}
\usepackage[table]{xcolor}
\usepackage[most]{tcolorbox}
\usepackage{colortbl}
\usepackage{xspace}
\usepackage{subcaption}
\usepackage{makecell}

\definecolor{mycolback}{RGB}{251,250,247}
\definecolor{mycolframe}{RGB}{172,201,235}
\newtcolorbox{mycommentboxenv}[1][]{%
  adjusted title={#1},
  colback=mycolback,
  colframe=mycolframe,
  boxrule=0.8pt,
  arc=4pt,
  left=6pt,
  right=6pt,
  top=1pt,
  bottom=1pt,
}
\newcounter{mycommentbox}[section] 
\newcommand{\mycommentbox}[3]{%
    \begin{mycommentboxenv}[#1]
    \refstepcounter{mycommentbox} 
    \label{#2}
    #3
    \end{mycommentboxenv}
}

\definecolor{lbcolor}{RGB}{13, 151, 175}
\newcommand{\method}{{MUSE}\xspace} 

\newcommand{\selectfull}{{Rank-Gain Utility Assessment}\xspace} 
\newcommand{\select}{{RGUA}\xspace} 
\newcommand{\score}{{trajectory utility score}\xspace} 

\newcommand{\contrastive}{{dual-level contrastive learning}\xspace} 

\newcommand{\genfull}{{Contrastive Trajectory Distillation}\xspace} 
\newcommand{\gen}{{ConDistill}\xspace} 

\newcommand{\recfull}{{Contrastive Reasoning Utilization}\xspace} 
\newcommand{\rec}{{ConUtil}\xspace} 

\newcommand{\training}{{Utility-Aware Contrastive Training}\xspace} 

\title{Multi-Teacher Reasoning Distillation for LLM Recommendation with Self-Exploration}

\author{%
   Liangbo Ning$^1$, Siyuan Liang$^2$, Yi Zhou$^1$, Yifeng Zhang$^1$, Qing Li$^1$, Wenqi Fan$^{1}$, Dacheng Tao$^2$ 
  \\ 
  $^1$ The Hong Kong Polytechnic University, Hong Kong SAR, China \\
  $^2$ Nanyang Technological University, Singapore  
  \\ 
  \texttt{\{BigLemon1123,wenqifan03\}@gmail.com}, \texttt{qing-prof.li@polyu.edu.hk},\\
  \texttt{\{echo-yi.zhou,yifeng.zhang\}@connect.polyu.hk},  
  \\
   \texttt{\{siyuan.liang,dacheng.tao\}@ntu.edu.sg}
}

\begin{document}

\maketitle


\begin{abstract}
Reasoning-augmented recommender systems (RecSys) enhance user preference modeling by generating reasoning trajectories over historical interactions before making recommendations. Knowledge distillation provides a practical paradigm for transferring such reasoning capabilities from powerful large language models to lightweight RecSys. However, existing methods typically rely on a single fixed teacher, remain constrained by the teacher knowledge ceiling, and indiscriminately distill teacher-generated trajectories without assessing their actual utility for downstream recommendation. To address these limitations, we propose MUSE (\underline{M}ulti-Teacher \underline{U}tility-guided \underline{S}election with Self-\underline{E}xploration), an adaptive reasoning distillation framework that jointly improves reasoning generation and reasoning utilization. MUSE first constructs a diverse reasoning pool using multiple teacher LLMs and progressively augments it via student self-exploration, enabling the student to leverage the strengths of different teachers and gradually explore reasoning strategies beyond teacher imitation. We then develop Rank-Gain Utility Assessment (RGUA), which evaluates each candidate according to the normalized full-catalog rank improvement of the ground-truth item and identifies the most and least recommendation-beneficial trajectories. Based on these utility-aware trajectory pairs, we introduce dual-level contrastive learning objectives that guide the student generator to prefer beneficial reasoning while training the recommendation encoder to more effectively exploit recommendation-relevant preference evidence. Extensive experiments on three real-world datasets demonstrate the effectiveness and robustness of the proposed framework.
\end{abstract}

\section{Introduction}

\begin{figure}
    \centering
    \includegraphics[width=0.65\linewidth]{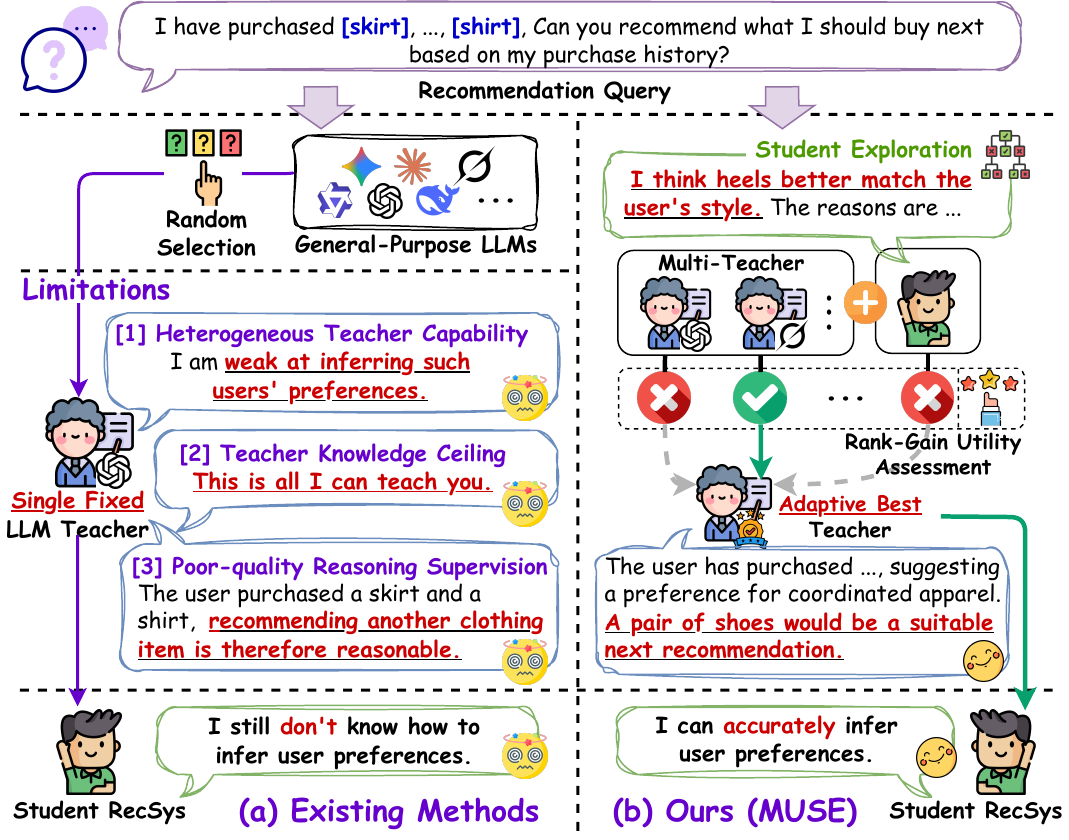}
    \caption{Illustration of the limitations of existing reasoning distillation methods and the proposed \method. 
    Existing reasoning distillation methods rely on a single fixed general-purpose teacher, leading to heterogeneous teacher capability, 
    teacher knowledge ceiling, and poor-quality reasoning supervision. 
    \method instead adaptively selects recommendation-beneficial reasoning from multiple teachers and progressively incorporates superior student-generated reasoning through self-exploration, enabling more effective reasoning distillation for recommendation. }
    \label{fig:intro}
    \vskip -0.2in
\end{figure}

Recommender systems (RecSys) have become a core component of Web platforms, helping users navigate information overload and discover relevant content, products, and services~\cite{zhao2024recommender,afsar2022reinforcement,hasan2025based,boeker2022empirical,covington2016deep,davidson2010youtube,wu2020mind,zheng2018drn,pan2024ads,theocharous2015ad}. 
With the rapid advancement of large language models (LLMs)~\cite{ren2026towards,chen2026mllms,liang2025explaining,liu2026survey,ho2026understanding,luo2026survey}, the recommendation paradigm has been fundamentally transformed, opening up a new era of LLM-empowered recommendation~\cite{zhao2024recommender,ning2024cheatagent,ning2026exploring,ning2026retrieval}. 
By representing users and items in natural language, LLM-empowered RecSys can leverage rich semantic knowledge and strong instruction-following capabilities to understand item descriptions and interpret user preferences, thereby delivering more accurate recommendations. 
More recently, the remarkable progress of reasoning-empowered LLMs (e.g., DeepSeek~\cite{ren2025deepseek,shao2024deepseekmath}) has motivated a further transition from language-based recommendation to \textbf{reasoning-augmented recommendation}. 
Rather than directly predicting the next item from an interaction sequence, a reasoning-augmented recommender system first generates an explicit or latent reasoning trajectory that analyzes the user's historical behaviors and infers the user's underlying preferences.
This trajectory is then used to produce more accurate recommendations.~\cite{you2026r,huang2026rerec,jiang2026diffureason,liang2026gr2}. 
Recent methods, such as R$^2$ec~\cite{you2026r} and ReRec~\cite{huang2026rerec}, have preliminarily explored this direction by encouraging LLM-based recommenders to reason over user histories before making predictions. 
Such preference reasoning provides a promising mechanism for capturing nuanced user interests while offering explicit evidence to support the generated recommendations.

Technologically speaking, existing approaches for acquiring reasoning capabilities for recommendations can be broadly divided into two paradigms: reinforcement learning (RL) and knowledge distillation. 
Reinforcement learning-based methods encourage a recommender to autonomously explore reasoning trajectories and optimize them according to carefully designed recommendation rewards~\cite{huang2026rerec,you2026r,gao2024end,zhang2025reinforced,tang2026think,ding2026towards}.
In principle, this trial-and-error process allows the model to discover reasoning strategies that are directly aligned with downstream ranking quality, instead of merely imitating human-written rationales. 
However, directly training a relatively small language model to acquire reliable reasoning capabilities remains difficult~\cite{guo2025deepseek,zhao2026self}. 
Owing to its limited model capacity, a small language model often struggles to generate sufficiently informative reasoning trajectories during the early stages of exploration, resulting in sparse or noisy reward signals, unstable optimization, and substantial computational overhead.
Knowledge distillation-based methods employ a powerful teacher LLM to generate preference reasoning trajectories for each user, which are subsequently used as supervision to train a smaller reasoning-augmented recommender system through standard supervised fine-tuning~\cite{wang2024can,tsai2024leveraging,wang2024rdrec,kim2025driven,qi2025knowledge}. 
Benefiting from the accurate supervision provided by the teacher model, knowledge distillation avoids relying solely on the student's initially weak exploration capability and typically yields a more stable optimization process than reinforcement learning from scratch. 
It is also complementary to reinforcement learning: a distilled model can serve as a strong initialization for subsequent RL-based post-training~\cite{zhang2025slow,hu2026adnanny,liang2026gr2,chen2026lasar}. 
Therefore, the favorable optimization stability and compatibility with further post-training collectively make knowledge distillation a practical paradigm for equipping lightweight recommender systems with explicit reasoning capabilities.

Despite this potential, as illustrated in Figure~\ref{fig:intro}, existing knowledge distillation-based methods still suffer from several limitations.
\textit{(1) Heterogeneous Teacher Capability.} 
Existing studies typically rely on a single fixed teacher to generate reasoning trajectories for subsequent distillation. 
However, due to differences in model architectures, pre-training corpora, and optimization objectives, LLMs often exhibit substantially different reasoning capabilities for recommendation tasks. 
A teacher that provides informative reasoning for one user, instance, or domain may generate inferior supervision for another. 
Consequently, reasoning trajectories produced by a single fixed teacher are unlikely to provide optimal supervision across diverse recommendation scenarios.
\textit{(2) Teacher Knowledge Ceiling.} 
General-purpose LLMs are primarily optimized for broad language understanding and generation rather than recommendation. 
As a result, they often lack sufficient recommendation-specific knowledge for accurately identifying users' latent preferences. 
Nevertheless, existing methods typically rely exclusively on reasoning trajectories generated by a fixed teacher throughout the entire distillation process. 
Consequently, the student model is fundamentally limited by the recommendation capability of its teacher, as it merely learns to imitate the teacher's reasoning behavior.
\textit{(3) Poor-quality Reasoning Supervision.} Existing methods usually treat all teacher-generated reasoning trajectories as equally reliable supervision without assessing their actual utility for downstream recommendation. 
A reasoning trajectory can be fluent, coherent, and seemingly persuasive while contributing little to recommendation performance, as it may emphasize preference signals that are irrelevant to retrieving the ground-truth item. 
Consequently, indiscriminately distilling such reasoning trajectories may fail to equip the downstream reasoning-augmented RecSys with reasoning capabilities that genuinely improve recommendation performance.

To address these limitations and enable recommenders to acquire more effective reasoning knowledge, we propose \method (\underline{M}ulti-Teacher \underline{U}tility-guided \underline{S}election with Self-\underline{E}xploration), an adaptive generation-then-selection framework that identifies and distills the reasoning trajectories most beneficial for recommendation.
Specifically, \method first introduces multiple teacher LLMs to construct a diverse reasoning pool, allowing different teachers to provide supervision for different users and thereby alleviating the issue of heterogeneous teacher capability. 
To overcome the knowledge ceiling of general-purpose teachers, \method further incorporates a progressive self-exploration strategy, in which the student progressively explores alternative reasoning trajectories as its recommendation capability improves. 
These self-generated trajectories are also included in the reasoning pool and evaluated together with teacher-generated candidates, enabling the student to gradually evolve beyond merely imitating its teachers. 
Then, we develop a \selectfull (\select) strategy that evaluates each candidate according to its contribution to downstream recommendation performance, enabling \method to select recommendation-beneficial reasoning while filtering out uninformative / poor-quality supervision. 
Finally, \method introduces the \contrastive objectives that encourage the RecSys to both accurately generate and effectively leverage recommendation-beneficial reasoning, leading to more accurate recommendations. 
The main contributions of this work are summarized as follows: 
\begin{itemize}[leftmargin=*]
    \item We propose \method, a multi-teacher reasoning distillation framework with progressive self-exploration, which alleviates heterogeneous teacher capability by using diverse teacher supervision and enables the student to move beyond the knowledge ceiling of general-purpose LLMs through progressive exploration.  
    \item We develop \selectfull (\select), which adaptively identifies recommendation-beneficial reasoning trajectories according to their downstream recommendation utility for knowledge distillation. Furthermore, we introduce a \contrastive objective that jointly improves reasoning generation and reasoning utilization, enabling the recommender to generate recommendation-beneficial reasoning and more effectively exploit it for accurate recommendations.
    \item We conduct extensive experiments on three real-world recommendation datasets. The results demonstrate the effectiveness of the proposed framework.
\end{itemize}

\section{Preliminaries}

\noindent
\textbf{Reasoning-Augmented Recommendation.}
Let $\mathcal{U}$ and $\mathcal{V}$ denote the user set and the complete item catalog, respectively. We consider a next-item recommendation dataset: $\mathcal{D}=\{(u_i,\mathcal{H}_i,y_i)\}_{i=1}^{N}$, 
where $u_i\in\mathcal{U}$ is a user, $\mathcal{H}_i=[(v_{i,1},a_{i,1}),\ldots,(v_{i,m_i},a_{i,m_i})]$ is the user's interaction history, $v_{i,j}\in\mathcal{V}$ and $a_{i,j}$ denote the interacted item and its textual attributes (e.g., item titles, descriptions), and $y_i\in\mathcal{V}$ is the ground-truth next item with attributes $a_{y_i}$. 
Generally, the user’s interaction history $\mathcal{H}_i$ is verbalized into a recommendation prompt $x_i$, which is then fed into the recommender system to generate the recommendation results. 
A reasoning-augmented RecSys usually consists of a reasoning trajectory generator $\mathcal{G}_{\theta}$ and a recommendation encoder $\mathcal{E}_{\phi}$~\cite{you2026r,huang2026rerec,zhang2025reinforced,gao2024end,wang2024can,lu2026think}. 
Conditioned on $x_i$, the generator first produces a textual / latent reasoning trajectory that summarizes the user's preferences:

\centerline{$r_i=\mathcal{G}_{\theta}(x_i).$}

\noindent
The recommendation encoder then maps the reasoning trajectory to a user representation and the textual attribute {$a_j$} of each item to the item representation, respectively: 

\centerline{$\mathbf{z}_i(r_i)=\mathcal{E}_{\phi}([x_i;r_i]), \quad
    \mathbf{q}_j=\mathcal{E}_{\phi}([v_j;a_j]).$}

\noindent
The relevance of item $v_j$ to user $u_i$ is measured by representation similarity (e.g., cosine similarity): 

\centerline{$s_{\phi}(u_i,v_j| r_i)
    =\operatorname{sim}(\mathbf{z}_i(r_i),\mathbf{q}_j).$}

\noindent
Items are sorted in descending order of $s_{\phi}(u_i,v_j| r_i)$, and the highest-ranked items are returned to the user as the final recommendation outcomes. 
It is worth noting that the recommendation module $\mathcal{E}_{\phi}$ can take different forms depending on the underlying architecture. 
In some reasoning-augmented RecSys, the reasoning generator itself is directly employed as the recommendation module~\cite{you2026r,huang2026rerec,zhang2025reinforced}, whereas other approaches adopt an additional lightweight encoder to encode the reasoning trajectories and candidate items for recommendation~\cite{gao2024end,wang2024can,kim2025driven}.

\noindent
\textbf{Reasoning Distillation.}
Recently, with the rapid advancement of LLM reasoning capabilities, increasing research attention has been devoted to transferring such reasoning abilities to recommender systems to enhance user preference modeling. 
Typically, reasoning distillation employs a powerful teacher LLM $\mathcal{T}$ to generate a trajectory for each training instance: $r_i^{\mathcal{T}}=\mathcal{T}(x_i).$
\noindent
Let $r_i^{\mathcal{T}}=[r_{i,1}^{\mathcal{T}},\ldots,r_{i,L_i}^{\mathcal{T}}]$ denote the teacher-generated reasoning trajectory for user $u_i$, where $L_i$ is its token length. 
The reasoning trajectory generator of the student reasoning-augmented RecSys is optimized using the standard autoregressive language modeling objective through teacher-forced supervised fine-tuning (SFT):

\centerline{$\mathcal{L}_{\mathrm{SFT}}
    =- \sum_{i=1}^{N}\sum_{t=1}^{L_i}
    \log \mathcal{G}_\theta\!\left(r_{i,t}^{\mathcal{T}}
    \mid x_i,r_{i,<t}^{\mathcal{T}}\right),$}

\noindent
where $r_{i,<t}^{\mathcal{T}}$ denotes the preceding reasoning tokens. 
After that, the student model learns to imitate the teacher-generated reasoning trajectories and produce preference reasoning conditioned on the user's historical interactions.

\begin{figure*}
    \centering
    \includegraphics[width=0.99\linewidth]{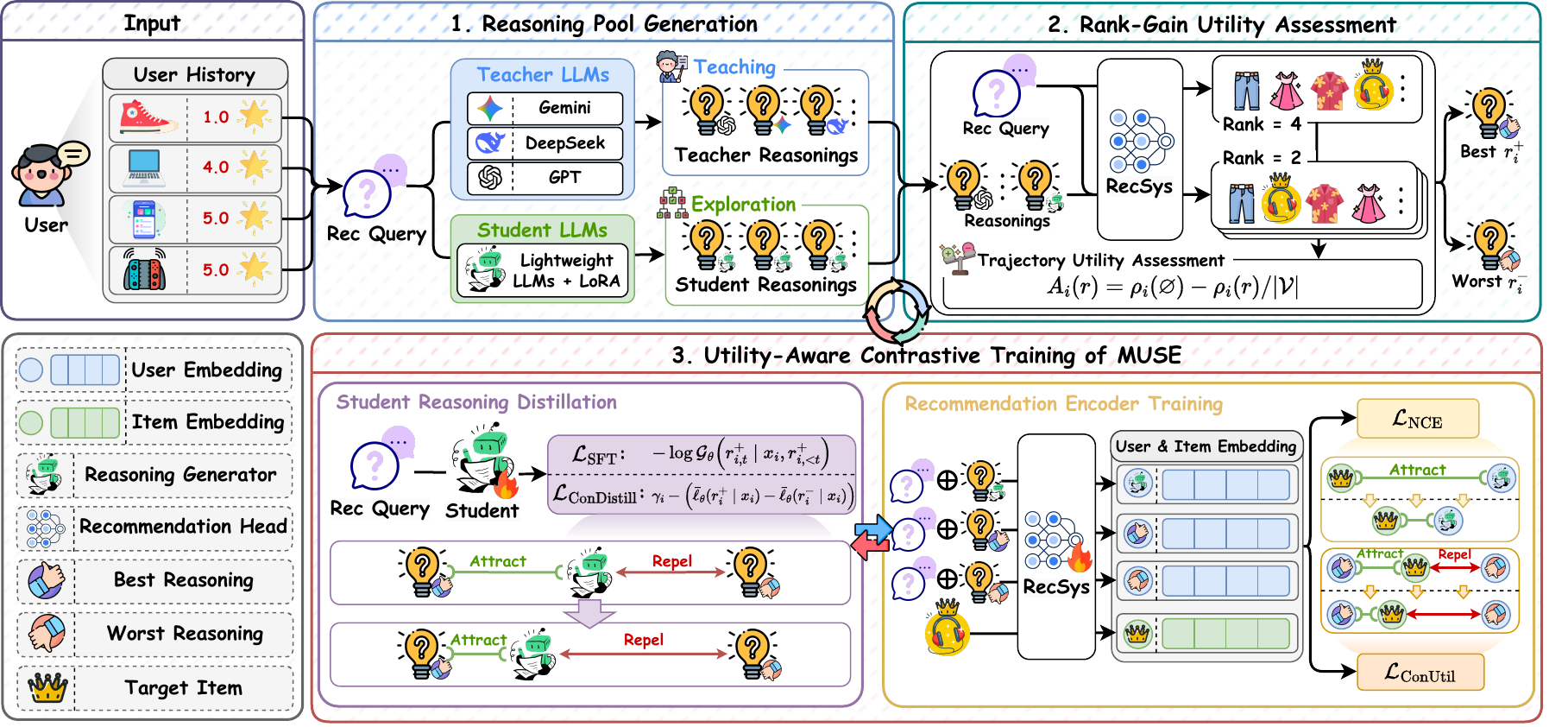}
    \caption{Overall architecture of \method. (1) Reasoning Pool Generation constructs diverse reasoning candidates from multiple teachers and progressive student self-exploration; (2) \selectfull identifies the most and least recommendation-beneficial reasoning trajectories according to their downstream ranking improvement; and (3) \training optimizes the reasoning generator and recommendation encoder using the \contrastive objectives.
    }
    \label{fig:framework}
    \vskip -0.2in
\end{figure*}

\section{Methodology}

\noindent
\textbf{Overview.}
To effectively distill reasoning knowledge from LLMs while overcoming the limitations of existing methods caused by relying on static general-purpose teachers, we propose \underline{M}ulti-Teacher \underline{U}tility-guided \underline{S}election with Self-\underline{E}xploration (\textbf{\method}).
As illustrated in Figure~\ref{fig:framework}, \method consists of three stages: Reasoning Pool Generation, \selectfull, and \training.
In the Reasoning Pool Generation stage, we introduce multiple teacher LLMs to generate diverse preference reasoning trajectories for each user, thereby avoiding reliance on a globally fixed teacher. 
We further propose a Progressive Self-Exploration strategy, in which the student progressively explores alternative reasoning trajectories as its capability improves and augments the candidate pool with these self-generated trajectories. 
In the \selectfull stage, each candidate reasoning trajectory is evaluated according to the improvement it brings to the full-catalog ranking of the ground-truth item relative to the reference recommendation without reasoning, allowing the most beneficial reasoning trajectory to be selected for supervised fine-tuning. 
Finally, during the \training stage, to more effectively exploit both high-quality and low-quality reasoning trajectories, we design the \contrastive objectives (i.e., \genfull and \recfull). 
The preferred and dispreferred reasoning provide comparative supervision for both the recommendation encoder and the student reasoning generator, encouraging the former to distinguish informative from misleading reasoning signals and the latter to generate recommendation-beneficial reasoning.

\subsection{Multi-Teacher Distillation with Self-Exploration}

\subsubsection{Multi-teacher reasoning pool construction.}
Different LLMs are trained with diverse architectures, optimization strategies, and pre-training corpora, resulting in substantially different reasoning capabilities for recommendation tasks. 
Arbitrarily selecting a single teacher largely overlooks whether it is genuinely well-suited for recommendation-specific reasoning, potentially limiting the effectiveness of the distilled student RecSys.
To address this issue, we propose a \emph{multi-teacher distillation} framework that leverages multiple teacher LLMs to generate diverse reasoning rationales for each user. 
By selecting the most appropriate reasoning trajectory for each user, the student recommender system can adaptively leverage the strengths of different teacher models, rather than being constrained by the capability of an arbitrarily selected single teacher.
Mathematically, we employ $M$ teacher LLMs, denoted by $\{\mathcal{T}_m\}_{m=1}^{M}$, and prompt each teacher with the same recommendation query to generate a preference reasoning trajectory for each user:
\begin{equation}
    r_i^{(m)}=\mathcal{T}_m(x_i), \qquad m=1,\ldots,M.
\end{equation}
The initial teacher candidate set for user $u_i$ is: 
\begin{equation}
    \mathcal{C}_{i}^{\mathrm{tea}}
    =\{r_i^{(1)},\ldots,r_i^{(M)}\}.
\end{equation}

\subsubsection{Progressive student self-exploration.}
Although we use multiple teachers, these teacher LLMs are primarily pretrained to acquire general-purpose language understanding and generation capabilities, and therefore inevitably suffer from limited recommendation-specific domain knowledge. 
Consequently, even the best teacher-generated reasoning trajectories may not always be optimal for recommendation tasks.
To overcome this limitation, we further introduce a self-exploration strategy. 
As training progresses and the reasoning capability of the student reasoning-augmented RecSys continuously improves, the student is encouraged not only to learn from the reasoning knowledge distilled from multiple teachers but also to autonomously explore alternative reasoning trajectories that may be more effective for recommendations. 
The self-generated reasoning trajectories are subsequently evaluated together with those produced by the teacher models, enabling the student to progressively discover reasoning strategies that are better aligned with recommendation objectives. 
In this way, the student evolves from merely imitating the teacher's reasoning to continuously improving through self-exploration, thereby alleviating the inherent limitation of teacher models with insufficient recommendation-specific knowledge.
Mathematically, let $\mathcal{G}_{\theta_e}$ denote the current student model at epoch $e$. We then sample $S_e$ reasoning trajectories based on its current parameters, defined as: $\tilde{r}_{i,e}^{(j)}\sim
    \mathcal{G}_{\theta_e}(r| x_i), j=1,\ldots,S_e,$
where $S_e$ denotes the sampling number.  
As the student model becomes increasingly capable during training, we progressively increase the number of reasoning trajectories sampled from the student to encourage more extensive self-exploration, formulated as: 
\begin{equation}
    S_e=
    \min \left(
        S_{\max},
        [ S_{\max} \cdot \frac{[e-e_0]_+}
        {E_{\mathrm{s}} - e_0}         ]
    \right),
    \label{eq:self_schedule}
\end{equation}
where $[e-e_0]_+=\max(0,e-e_0)$, $e_0=2$ is the warm-up epoch, $S_{\max}$ is the maximum number of student reasoning candidates, and $E_{\mathrm{s}}$ denotes the epoch at which the number of sampled trajectories reaches its maximum value $S_{\max}$.  
Finally, the complete candidate reasoning pool $\mathcal{C}_{i}^{(e)}$ becomes: 
\begin{equation}
    \mathcal{C}_{i}^{(e)}
    =\mathcal{C}_{i}^{\mathrm{tea}}
    \cup\{\tilde{r}_{i,e}^{(1)},\ldots,
    \tilde{r}_{i,e}^{(S_e)}\}.
    \label{eq:candidate_pool}
\end{equation}

\subsection{\selectfull (\select)}

After constructing a diverse candidate pool through multiple teachers and self-exploration, the remaining challenge is to identify high-quality reasoning trajectories without human annotations. 
We argue that useful reasoning should consistently improve recommendation quality by helping the model better identify the user's true preference. Therefore, instead of evaluating reasoning in isolation, we can directly measure its contribution to the downstream recommendation objective through the ranking improvement of the ground-truth item. 
Based on this intuition, we propose a \selectfull strategy.
Specifically, for each training instance, we first encode the recommendation prompt without appending any reasoning trajectory and compute the rank of the ground-truth item $y_i$ over the complete catalog:
\begin{equation}
    \rho_i(\varnothing)
    ={ \sum_{v_j\in\mathcal{V}}}~
    \mathbb{I}\!\left[
        s_{\phi}(u_i,v_j\mid\varnothing)
        >s_{\phi}(u_i,y_i\mid\varnothing)
    \right].
    \label{eq:rank_without}
\end{equation}
For a candidate $r\in\mathcal{C}_{i}^{(e)}$, we append it to the same prompt and recompute the rank:
\begin{equation}
    \rho_i(r)
    ={ \sum_{v_j\in\mathcal{V}}}~
    \mathbb{I}\!\left[
        s_{\phi}(u_i,v_j\mid r)
        >s_{\phi}(u_i,y_i\mid r)
    \right].
    \label{eq:rank_with}
\end{equation}
A lower value of $\rho_i(r)$ indicates that the candidate places the target closer to the top of the ranking list. Then, we define the \score as the normalized rank improvement:
\begin{equation}
    A_i(r)=\frac{\rho_i(\varnothing)-\rho_i(r)}{|\mathcal{V}|},
    \label{eq:advantage}
\end{equation}
\noindent
where $A_i(r)>0$ means that reasoning improves the target rank 
and $A_i(r)<0$ indicates that it worsens the target-item rank relative to the no-reasoning input.
Normalization by the catalog size constrains the utility score to $(-1,1)$ and makes the scale comparable across datasets with different catalog sizes.
Intuitively, trajectories that yield larger ranking improvements better capture the user's latent preferences and are therefore considered more beneficial for the downstream recommendation task. 
Therefore, MUSE selects the trajectories with the highest and lowest assessed utility as the preferred and dispreferred candidates, respectively:
\begin{equation}
    \begin{aligned}
        r_i^{+}=\arg\max_{r\in\mathcal{C}_{i}^{(e)}}& A_i(r),  \quad   r_i^{-}=\arg\min_{r\in\mathcal{C}_{i}^{(e)}} A_i(r), \\
    \Delta_i^{A}&=A_i(r_i^{+})-A_i(r_i^{-}).
    \label{eq:best_worst}
    \end{aligned}
\end{equation}
The trajectory $r_i^{+}$ serves as the primary distillation target, while the pair $(r_i^{+},r_i^{-})$ captures which reasoning should be preferred and how strongly it should be preferred.

\subsection{\training of \method}

\method consists of two trainable components: a reasoning generator that learns to generate recommendation-beneficial preference reasoning, and a recommendation encoder that learns to leverage the generated reasoning to produce the final recommendation. 
During training, to more effectively exploit reasoning trajectories of varying quality/utility, we introduce the \contrastive objectives (i.e., \genfull and \recfull) for these two key components.
Next, we present the training details for each component.

\subsubsection{Student reasoning distillation.}
The primary objective of this stage is to identify suitable reasoning trajectories that serve as supervision targets to guide the student's distillation learning.
Since \selectfull already explicitly evaluates the recommendation utility of different reasoning trajectories, it naturally provides effective supervision signals for training the student reasoning generator. 
Specifically, for each training instance, we select the candidate trajectory $r_i^{+}$ with the highest \score as the distillation target, since it most accurately captures the user's true preferences and thus contributes the most to the downstream recommendation objective. 
The student generator is then optimized using the standard autoregressive negative log-likelihood objective:
\begin{equation}
    \mathcal{L}_{\mathrm{SFT}}
    =-
    \sum_{i\in\mathcal{B}}\sum_{t=1}^{L_i^{+}}
    \log \mathcal{G}_{\theta}\!\left(
        r_{i,t}^{+}\mid x_i,r_{i,<t}^{+}
    \right).
    \label{eq:agred_sft}
\end{equation}

\noindent However, this standard knowledge distillation only encourages the student to imitate the selected high-quality reasoning trajectory $r_i^{+}$, but provides no explicit supervision on which reasoning trajectories should be avoided.
Fortunately, the proposed \select enables us to readily identify not only the most but also the least recommendation-beneficial reasoning trajectories according to their utility for downstream recommendation. 
Based on these reasoning pairs, we further introduce a \textbf{\genfull (\gen)} objective that encourages the student to assign a higher likelihood to the most beneficial reasoning trajectory than to the least beneficial one, formulated as: 
\begin{equation}
    \mathcal{L}_{\mathrm{\gen}}
    =\frac{1}{|\mathcal{B}|}
    \sum_{i\in\mathcal{B}}
    \left[
        \gamma_i
        -\left(
        \bar{\ell}_{\theta}(r_i^{+}\mid x_i)
        -\bar{\ell}_{\theta}(r_i^{-}\mid x_i)
        \right)
    \right]_+,
    \label{eq:smpa}
\end{equation}
where $\bar{\ell}_{\theta}(r| x_i)=\frac{1}{|r|}\sum_{t=1}^{|r|} \log \mathcal{G}_{\theta}(r_t| x_i,r_{<t})$ is the length-normalized log-likelihood of a trajectory and $\mathcal{B}$ is the current batch data. Length normalization prevents longer trajectories from being systematically penalized by accumulated token log-likelihood. 
$\gamma_i=[\Delta_i^{A}]_+$ is the desired likelihood-gap margin for instance $i$. 
A pair with a large trajectory utility gap imposes a stronger separation requirement, whereas similar-quality trajectories induce a smaller margin, thereby enabling adaptive margin adjustment. 
This contrastive optimization objective explicitly enlarges the likelihood gap between recommendation-beneficial and recommendation-detrimental reasoning trajectories, thereby enabling the student reasoning generator to better discriminate high-quality reasoning during training.
Finally, combining the autoregressive SFT loss with the proposed \gen objective, the overall training objective of the student generator is defined as: 
\begin{equation}
    \mathcal{L}_{\mathrm{gen}}
    =\mathcal{L}_{\mathrm{SFT}}
    +\lambda_{\mathrm{\gen}}\mathcal{L}_{\mathrm{\gen}},
    \label{eq:gen_total}
\end{equation}
where $\lambda_{\mathrm{\gen}}=0.5$ balances the contributions of different terms.

\subsubsection{Recommendation encoder training.}
While the preceding stage trains the student generator to produce recommendation-beneficial reasoning trajectories, accurate recommendations further require the recommendation encoder to effectively utilize the generated reasoning for user preference modeling. 
Therefore, the primary objective of this stage is to train the recommendation encoder to align the reasoning-enhanced user representation with the target item representation. 
At inference time, generating reasoning trajectories with multiple teacher LLMs is impractical due to the stringent latency requirement of online recommendation. 
Consequently, the recommendation encoder must consume reasoning generated solely by the student model during inference. 
Training the recommendation encoder exclusively with teacher-generated reasoning would therefore create a train--test discrepancy.
To maintain consistency, the current student first generates a trajectory $r_i^{\mathrm{cur}}$ for each training user. 
Then, the user and the corresponding target item representations are denoted by: $\mathbf{z}_i^{\mathrm{cur}}=\mathcal{E}_{\phi}([x_i;r_i^{\mathrm{cur}}]), \mathbf{q}_i^{+}=\mathcal{E}_{\phi}([{y_i}; a_{y_i}])$.
\noindent
For a mini-batch $\mathcal{B}$, we first optimize the recommendation encoder using the typical in-batch softmax~\cite{you2026r,oord2018representation}:
\begin{equation}
    \mathcal{L}_{\mathrm{NCE}}
    =-\frac{1}{|\mathcal{B}|}
    \sum_{i\in\mathcal{B}}
    \log
    \frac{
        \exp(s_{\phi}(u_i, y_i|r_i^{\text{cur}})/\tau_r)
    }{
        \sum_{j\in\mathcal{B}}
        \exp(s_{\phi}(u_i, y_j|r_i^{\text{cur}})/\tau_r)
    },
    \label{eq:infonce}
\end{equation}
where $s_{\phi}(u_i, y_i|r_i^{\text{cur}}) = \operatorname{sim}(\mathbf{z}_i^{\mathrm{cur}}, \mathbf{q}_i^{+})$, $\tau_r=0.1$ is the temperature, and the target items of other users serve as negatives.

However, the standard recommendation objective provides only outcome-level supervision: it requires the resulting user representation to align with the target item, but does not constrain the encoder to achieve this alignment by exploiting recommendation-relevant reasoning signals.
Since the interaction history itself can provide strong predictive cues, the encoder may reduce the recommendation loss by relying primarily on historical interactions, leaving the preference information conveyed by the reasoning trajectory insufficiently utilized.
To provide additional supervision across alternative reasoning trajectories, we introduce \textbf{\recfull (\rec)}, which exploits the preferred and dispreferred reasoning trajectories to provide fine-grained supervision for reasoning-aware representation learning. 
Specifically, \rec contrasts the user representations induced by $r_i^{+}$ and $r_i^{-}$ while keeping the interaction history and target item unchanged.
This controlled comparison complements the standard outcome-level supervision by explicitly revealing how reasoning quality should influence target-item matching.
As a result, $r_i^{+}$ is encouraged to induce a user representation that is more strongly aligned with the target item than $r_i^{-}$, enabling the recommendation encoder to emphasize recommendation-beneficial preference evidence while attenuating irrelevant or misleading reasoning signals.
Formally, the user representations based on different reasonings are defined as: 
$\mathbf{z}_i^{+}=\mathcal{E}_{\phi}([x_i;r_i^{+}]),
    \mathbf{z}_i^{-}=\mathcal{E}_{\phi}([x_i;r_i^{-}]).$
\rec penalizes cases where the target item representation $\mathbf{q}_i^{+}=\mathcal{E}_{\phi}([{y_i}; a_{y_i}])$ is more similar to the user representation induced by $r_i^{-}$ than to that induced by $r_i^{+}$:
\begin{equation}
    \mathcal{L}_{\mathrm{\rec}}
    =\frac{1}{|\mathcal{B}|}
    \sum_{i\in\mathcal{B}}
    \left[
    s_{\phi}(u_i, y_i | r^-) - s_{\phi}(u_i, y_i | r^+)
    \right]_+, 
    \label{eq:rcr}
\end{equation}
where $s_{\phi}(u_i, y_i | r^-) = \operatorname{sim}(\mathbf{z}_i^{-},\mathbf{q}_i^{+})$ and $s_{\phi}(u_i, y_i | r^+)=\operatorname{sim}(\mathbf{z}_i^{+},\mathbf{q}_i^{+})$.
In this way, \rec encourages the recommendation encoder to explicitly exploit recommendation-relevant preference evidence.
Finally, the overall objective for the recommendation encoder is:
\begin{equation}
    \mathcal{L}_{\mathrm{rec}}
    =\mathcal{L}_{\mathrm{NCE}}
    +\lambda_{\mathrm{\rec}}\mathcal{L}_{\mathrm{\rec}},
    \label{eq:rec_total}
\end{equation}
where $\lambda_{\mathrm{\rec}}=0.5$ controls the contribution of different terms.

\subsection{Inference}

\method requires neither teacher LLMs nor trajectory utility estimation during inference. Given a new user's prompt $x_i$, the well-trained student generator first produces a preference reasoning trajectory: $\hat{r}_i=
    \mathcal{G}_{\theta}(x_i).$
The recommendation encoder then computes the user representation $\hat{\mathbf{z}}_i=\mathcal{E}_{\phi}([x_i;\hat{r}_i])$
and compares it with the representations of all catalog items using cosine similarity. 
The final recommendation list is defined as: 
\begin{equation}
    \widehat{\mathcal{R}}_i^{K}
    =\operatorname{TopK}_{v_j\in
    \mathcal{V}}
    s_{\phi}(u_i,v_j|\hat{r}_i)
    =\operatorname{TopK}_{v_j\in
    \mathcal{V}}
    \operatorname{sim}(\hat{\mathbf{z}}_i,\mathbf{q}_j).
    \label{eq:final_rank}
\end{equation}
The resulting pipeline requires neither costly teacher calls nor candidate evaluation at test time, while retaining the recommendation-beneficial reasoning discovered from multiple teachers and the student's own exploration.
The complete training and inference procedures are summarized in \textbf{Algorithm}~\ref{al:framework}.

\section{Experiments}

\subsection{Experimental Details}

\noindent
\textit{\textbf{4.1.1 Datasets.}}
Following prior studies~\cite{you2026r,bao2024decoding,bao2025bi}, we conduct experiments on three publicly available real-world datasets from Amazon, namely \textbf{CDs and Vinyl} (\textbf{CDs}), \textbf{Video Games} (\textbf{Games}), and \textbf{Musical Instruments} (\textbf{Instruments}). 
We adopt the widely used leave-one-out evaluation protocol~\cite{qu2024tokenrec}.
Detailed dataset statistics are summarized in Table~\ref{tab:dataset}.
To evaluate recommendation performance, we employ two widely used ranking metrics, \textbf{Hit Rate (H@K)} and \textbf{Normalized Discounted Cumulative Gain (N@K)}, with $K\in \{5,10,20\}$. 
Following common practice, we adopt the full-ranking evaluation protocol, where each user's ranking metrics are computed over the entire candidate item set.
\begin{table}[htbp]
\vskip -0.15in
  \centering
  \caption{Dataset Statistics.}
  \scalebox{0.8}{
    \begin{tabular}{c|ccc}
    \toprule
    \textbf{Dataset} & \textbf{Users} & \textbf{Items} & \textbf{Interactions} \\
    \midrule
    \textbf{Video Games} & 29,230 & 10,144 &  63,502 \\
    \textbf{CDs and Vinyl} & 7,701 & 12,024 &  13,435 \\
    \textbf{Musical Instruments} & 15,656 & 10,320 &  34,373 \\
    \bottomrule
    \end{tabular}%
    }
  \label{tab:dataset}%
  \vskip -0.15in
\end{table}%

\noindent
\textit{\textbf{4.1.2 Baselines.}}
To comprehensively evaluate the effectiveness of the proposed method, we compare it against a diverse set of competitive baselines from different categories, including traditional sequential RecSys (\textbf{GRU4Rec}~\cite{hidasi2015session}, \textbf{Caser}~\cite{tang2018personalized}, \textbf{SASRec}~\cite{kang2018self}), LLM-based RecSys (\textbf{TIGER}~\cite{rajput2023recommender}, \textbf{BigRec}~\cite{bao2025bi}, and \textbf{D$^3$}~\cite{bao2024decoding}), and reasoning-augmented RecSys (\textbf{RecSAVER}~\cite{tsai2024leveraging}, \textbf{LangPTune}~\cite{gao2024end}, and \textbf{R$^2$ec}~\cite{you2026r}). 
More detailed descriptions of these baselines are provided in \textbf{Appendix~\ref{appendix:baselines}}.

\noindent
\textit{\textbf{4.1.3 Implementation.}}
All baselines and \method are implemented in PyTorch.
We instantiate \method with two student reasoning generators, \texttt{Gemma-2-2B}~\cite{team2024gemma} and \texttt{Qwen2.5-3B}~\cite{qwen2.5}, together with a fully fine-tuned recommendation encoder, \texttt{GTE-large-en-v1.5}~\cite{zhang2024mgte,li2023towards}, which produces 1024-dimensional CLS-pooled embeddings with a maximum input length of 2,048 tokens.
For more implementation details of \method and the prompts used, please refer to \textbf{Appendix~\ref{appendix:implementation}} and \textbf{Appendix~\ref{appendix:used_prompt}}.

\subsection{Overall Performance}\label{exp:overall_performance}
Table~\ref{tab:main_result_cds}, Table~\ref{tab:main_result_music}, and Table~\ref{tab:main_result_video} (\textbf{Appendix~\ref{appendix:main_experiments}}) report the overall recommendation performance of different methods. Several important observations can be drawn. 
    \textit{(1) First}, incorporating explicit preference reasoning consistently improves recommendation performance. Across almost all datasets and evaluation metrics, reasoning-augmented RecSys, including RecSAVER, LangPTune, R$^2$ec, and the proposed \method, consistently outperform both traditional sequential recommenders (e.g., Caser and GRU4Rec) and LLM-based RecSys baselines without explicit reasoning (e.g., D$^3$ and BigRec). 
    These results demonstrate that explicitly modeling users' latent preferences through reasoning provides more informative semantic signals for user preference modeling, leading to substantially better recommendation performance.    
    \textit{(2) Second}, \method consistently outperforms RL-based methods, including LangPTune and R$^2$ec, despite relying solely on reasoning distillation during training. 
    This result suggests that carefully selecting high-quality reasoning trajectories can be more effective than directly optimizing reasoning through reinforcement learning. 
    Moreover, reasoning distillation is naturally complementary to RL-based post-training, allowing the distilled student model to serve as a stronger initialization for subsequent reinforcement learning. 
    Therefore, the proposed framework not only achieves superior recommendation performance by itself but also provides a more effective foundation for further reasoning optimization.
    \textit{(3) Third}, the proposed \method consistently achieves the best overall performance among all reasoning-based recommendation methods. 
    Compared with the distillation-based baseline RecSAVER, which relies on a single fixed teacher for distillation, \method constructs a diverse reasoning pool using multiple teachers and progressive self-exploration, and further selects the most recommendation-beneficial reasoning trajectories through the proposed \selectfull. 
    This dynamic reasoning distillation paradigm consistently yields more effective supervision, resulting in substantial improvements across almost all datasets and metrics.
    Due to space limitations, additional case studies are provided in \textbf{Appendix~\ref{appendix:reasoning_cases}}.
    

\begin{table*}[htbp]
  \centering
  \caption{The overall performance of baselines and \method (Dataset: CDs and Vinyl).}
  \vskip -0.1in
  \scalebox{0.7}{
    \begin{tabular}{|c|c|c|c|c|c|c|}
    \toprule
    \textbf{Datasets} & \multicolumn{6}{c|}{\textbf{CDs and Vinyl}} \\
    \midrule
    \textbf{Methods} & \textbf{N@5} & \textbf{N@10} & \textbf{N@20} & \textbf{H@5} & \textbf{H@10} & \textbf{H@20} \\
    \midrule
    Caser & 0.0034±0.0003 & 0.0046±0.0008 & 0.0061±0.0005 & 0.0068±0.0008 & 0.0105±0.0021 & 0.0166±0.0011 \\
    GRU4Rec & 0.0036±0.0013 & 0.0049±0.0016 & 0.0068±0.0021 & 0.0059±0.0022 & 0.0102±0.0033 & 0.0178±0.0049 \\
    SASRec & 0.0051±0.0009 & 0.0060±0.0009 & 0.0070±0.0006 & 0.0084±0.0007 & 0.0114±0.0008 & 0.0154±0.0014 \\
    TIGER & 0.0023±0.0009 & 0.0034±0.0009 & 0.0041±0.0006 & 0.0033±0.0011 & 0.0069±0.0011 & 0.0095±0.0003 \\
    \midrule
    \rowcolor[rgb]{ .929,  .929,  .929} \multicolumn{7}{|c}{\textbf{Qwen2.5-3B}} \\
    \midrule
    D$^3$    & 0.0030±0.0024 & 0.0048±0.0028 & 0.0070±0.0033 & 0.0054±0.0041 & 0.0110±0.0053 & 0.0196±0.0073 \\
    BigRec & 0.0031±0.0009 & 0.0033±0.0011 & 0.0037±0.0011 & 0.0040±0.0012 & 0.0048±0.0017 & 0.0064±0.0020 \\
    RecSAVER & 0.0138±0.0014 & 0.0175±0.0032 & 0.0215±0.0023 & 0.0201±0.0051 & 0.0315±0.0105 & 0.0475±0.0069 \\
    LangPTune & 0.0231±0.0011 & 0.0289±0.0006 & 0.0338±0.0010 & 0.0349±0.0008 & 0.0531±0.0011 & 0.0729±0.0007 \\
    R$^2$ec  & \cellcolor[rgb]{ .851,  .882,  .949}\underline{0.0243±0.0006} & \cellcolor[rgb]{ .851,  .882,  .949}\underline{0.0302±0.0002} & \cellcolor[rgb]{ .851,  .882,  .949}\underline{0.0352±0.0005} & \cellcolor[rgb]{ .851,  .882,  .949}\underline{0.0362±0.0011} & \cellcolor[rgb]{ .851,  .882,  .949}\underline{0.0545±0.0008} & \cellcolor[rgb]{ .851,  .882,  .949}\underline{0.0743±0.0012} \\
    \textbf{\method}  & \cellcolor[rgb]{ .965,  .792,  .745}\textbf{0.0295±0.0009} & \cellcolor[rgb]{ .965,  .792,  .745}\textbf{0.0348±0.0017} & \cellcolor[rgb]{ .965,  .792,  .745}\textbf{0.0404±0.0010} & \cellcolor[rgb]{ .965,  .792,  .745}\textbf{0.0421±0.0019} & \cellcolor[rgb]{ .965,  .792,  .745}\textbf{0.0584±0.0041} & \cellcolor[rgb]{ .965,  .792,  .745}\textbf{0.0804±0.0018} \\
    \midrule
    \rowcolor[rgb]{ .929,  .929,  .929} \multicolumn{7}{|c}{\textbf{Gemma-2-2B}} \\
    \midrule
    D$^3$    & 0.0015±0.0011 & 0.0019±0.0007 & 0.0031±0.0008 & 0.0026±0.0015 & 0.0040±0.0011 & 0.0086±0.0004 \\
    BigRec & 0.0026±0.0004 & 0.0028±0.0003 & 0.0039±0.0004 & 0.0039±0.0003 & 0.0058±0.0006 & 0.0093±0.0005 \\
    RecSAVER & 0.0154±0.0030 & 0.0196±0.0029 & 0.0243±0.0029 & 0.0212±0.0053 & 0.0349±0.0051 & 0.0533±0.0050 \\
    LangPTune & 0.0204±0.0026 & 0.0254±0.0025 & 0.0302±0.0028 & 0.0285±0.0039 & 0.0442±0.0037 & 0.0631±0.0046 \\
    R$^2$ec  & \cellcolor[rgb]{ .851,  .882,  .949}\underline{0.0234±0.0020} & \cellcolor[rgb]{ .851,  .882,  .949}\underline{0.0282±0.0010} & \cellcolor[rgb]{ .851,  .882,  .949}\underline{0.0355±0.0008} & \cellcolor[rgb]{ .851,  .882,  .949}\underline{0.0355±0.0039} & \cellcolor[rgb]{ .851,  .882,  .949}\underline{0.0503±0.0016} & \cellcolor[rgb]{ .851,  .882,  .949}\underline{0.0795±0.0011} \\
    \textbf{\method}  & \cellcolor[rgb]{ .965,  .792,  .745}\textbf{0.0273±0.0029} & \cellcolor[rgb]{ .965,  .792,  .745}\textbf{0.0332±0.0035} & \cellcolor[rgb]{ .965,  .792,  .745}\textbf{0.0394±0.0049} & \cellcolor[rgb]{ .965,  .792,  .745}\textbf{0.0379±0.0039} & \cellcolor[rgb]{ .965,  .792,  .745}\textbf{0.0561±0.0064} & \cellcolor[rgb]{ .965,  .792,  .745}\textbf{0.0806±0.0122} \\
    \bottomrule
    \multicolumn{7}{l}{\colorbox[rgb]{ .965,  .792,  .745}{\textbf{Red}} and \colorbox[rgb]{ .851,  .882,  .949}{\underline{Blue}} denote the best and second-best results, respectively.}
    \end{tabular}%
    }
  \label{tab:main_result_cds}%
\end{table*}%

\begin{table}[htbp]
  \centering
  \caption{The overall performance of baselines and \method (Dataset: Musical Instruments).}
  \scalebox{0.7}{
    \begin{tabular}{|c|c|c|c|c|c|c|}
    \toprule
    \textbf{Datasets} & \multicolumn{6}{c|}{\textbf{Musical Instruments}} \\
    \midrule
    \textbf{Methods} & \textbf{N@5} & \textbf{N@10} & \textbf{N@20} & \textbf{H@5} & \textbf{H@10} & \textbf{H@20} \\
    \midrule
    Caser & 0.0087±0.0022 & 0.0134±0.0025 & 0.0191±0.0018 & 0.0169±0.0043 & 0.0317±0.0055 & 0.0543±0.0034 \\
    GRU4Rec & 0.0090±0.0007 & 0.0117±0.0010 & 0.0152±0.0017 & 0.0143±0.0013 & 0.0225±0.0022 & 0.0367±0.0048 \\
    SASRec & 0.0058±0.0010 & 0.0097±0.0004 & 0.0133±0.0012 & 0.0099±0.0013 & 0.0219±0.0018 & 0.0362±0.0058 \\
    TIGER & 0.0139±0.0038 & 0.0173±0.0034 & 0.0218±0.0022 & 0.0208±0.0050 & 0.0313±0.0041 & 0.0491±0.0011 \\
    \midrule
    \rowcolor[rgb]{ .929,  .929,  .929} \multicolumn{7}{|c|}{\textbf{Qwen2.5-3B}} \\
    \midrule
    D$^3$    & 0.0089±0.0024 & 0.0124±0.0036 & 0.0152±0.0051 & 0.0170±0.0044 & 0.0273±0.0081 & 0.0386±0.0141 \\
    BigRec & 0.0038±0.0006 & 0.0049±0.0007 & 0.0059±0.0003 & 0.0061±0.0003 & 0.0087±0.0004 & 0.0123±0.0015 \\
    RecSAVER & 0.0105±0.0025 & 0.0144±0.0021 & 0.0189±0.0022 & 0.0156±0.0031 & 0.0279±0.0022 & 0.0456±0.0024 \\
    LangPTune & \cellcolor[rgb]{ .851,  .882,  .949}\underline{0.0159±0.0009} & \cellcolor[rgb]{ .851,  .882,  .949}\underline{0.0209±0.0008} & 0.0258±0.0009 & \cellcolor[rgb]{ .851,  .882,  .949}\underline{0.0241±0.0019} & 0.0397±0.0025 & 0.0590±0.0026 \\
    R$^2$ec  & 0.0133±0.0005 & 0.0195±0.0012 & \cellcolor[rgb]{ .851,  .882,  .949}\underline{0.0274±0.0006} & 0.0206±0.0009 & \cellcolor[rgb]{ .851,  .882,  .949}\underline{0.0402±0.0029} & \cellcolor[rgb]{ .851,  .882,  .949}\underline{0.0713±0.0006} \\
    \textbf{\method}  & \cellcolor[rgb]{ .965,  .792,  .745}\textbf{0.0194±0.0008} & \cellcolor[rgb]{ .965,  .792,  .745}\textbf{0.0255±0.0005} & \cellcolor[rgb]{ .965,  .792,  .745}\textbf{0.0323±0.0013} & \cellcolor[rgb]{ .965,  .792,  .745}\textbf{0.0290±0.0011} & \cellcolor[rgb]{ .965,  .792,  .745}\textbf{0.0480±0.0008} & \cellcolor[rgb]{ .965,  .792,  .745}\textbf{0.0755±0.0030} \\
    \midrule
    \rowcolor[rgb]{ .929,  .929,  .929} \multicolumn{7}{|c|}{\textbf{Gemma-2-2B}} \\
    \midrule
    D$^3$    & 0.0090±0.0033 & 0.0141±0.0031 & 0.0201±0.0009 & 0.0148±0.0058 & 0.0306±0.0056 & 0.0547±0.0080 \\
    BigRec & 0.0010±0.0005 & 0.0017±0.0005 & 0.0029±0.0006 & 0.0011±0.0005 & 0.0032±0.0005 & 0.0075±0.0010 \\
    RecSAVER & 0.0103±0.0008 & 0.0138±0.0003 & 0.0186±0.0003 & 0.0157±0.0020 & 0.0270±0.0027 & 0.0458±0.0041 \\
    LangPTune & 0.0125±0.0020 & 0.0177±0.0018 & 0.0221±0.0011 & 0.0190±0.0034 & 0.0353±0.0029 & 0.0529±0.0026 \\
    R$^2$ec  & \cellcolor[rgb]{ .851,  .882,  .949}\underline{0.0174±0.0003} & \cellcolor[rgb]{ .851,  .882,  .949}\underline{0.0231±0.0006} & \cellcolor[rgb]{ .851,  .882,  .949}\underline{0.0294±0.0016} & \cellcolor[rgb]{ .851,  .882,  .949}\underline{0.0259±0.0014} & \cellcolor[rgb]{ .851,  .882,  .949}\underline{0.0436±0.0021} & \cellcolor[rgb]{ .851,  .882,  .949}\underline{0.0686±0.0063} \\
    \textbf{\method}  & \cellcolor[rgb]{ .965,  .792,  .745}\textbf{0.0195±0.0009} & \cellcolor[rgb]{ .965,  .792,  .745}\textbf{0.0240±0.0011} & \cellcolor[rgb]{ .965,  .792,  .745}\textbf{0.0317±0.0011} & \cellcolor[rgb]{ .965,  .792,  .745}\textbf{0.0295±0.0009} & \cellcolor[rgb]{ .965,  .792,  .745}\textbf{0.0448±0.0007} & \cellcolor[rgb]{ .965,  .792,  .745}\textbf{0.0742±0.0011} \\
    \bottomrule
    \multicolumn{7}{l}{\colorbox[rgb]{ .965,  .792,  .745}{\textbf{Red}} and \colorbox[rgb]{ .851,  .882,  .949}{\underline{Blue}} denote the best and second-best results, respectively.}
    \end{tabular}%
    }
  \label{tab:main_result_music}%
\end{table}%

\subsection{Ablation Study}\label{exp:ablation_study}
We first compare \method with three single-teacher variants, where the reasoning trajectories are generated solely by \textit{GPT}, \textit{Gemini}, or \textit{DeepSeek}, respectively, to evaluate the effectiveness of the proposed multi-teacher distillation strategy. 
We further investigate the contribution of each proposed component. 
\textit{\method w/o SE} disables the proposed progressive self-exploration strategy, leaving only teacher-generated reasoning trajectories for distillation. 
\textit{\method-Rand} replaces the proposed \select by randomly selecting a reasoning trajectory from the teacher- and self-generated candidate pool for supervision. 
Finally, \textit{\method w/o \gen} removes the proposed \gen objective for the reasoning generator, while \textit{\method w/o \rec} removes the proposed \rec objective for the recommendation encoder. 
Table~\ref{tab:ablation_study_cds}, Table~\ref{tab:ablation_study_music}, and Table~\ref{tab:ablation_study_video} (\textbf{Appendix~\ref{appendix:ablation_study}}) report the experimental results. Several important observations can be made.

First, multi-teacher distillation consistently outperforms all single-teacher variants across all datasets. 
This demonstrates that no single teacher can consistently provide the most recommendation-beneficial reasoning for every user. By leveraging diverse reasoning trajectories from multiple teachers and selecting the most effective one through the proposed \select, \method provides higher-quality supervision than relying on any individual teacher alone.
Second, each proposed component contributes positively to the overall performance. 
Disabling progressive self-exploration (\textit{\method w/o SE}) results in inferior performance, suggesting that allowing the student to continuously discover recommendation-beneficial reasoning beyond the fixed teacher pool effectively alleviates the teacher knowledge ceiling. 
Replacing \selectfull with random trajectory selection (\textit{\method-Rand}) leads to a noticeable performance degradation, validating the importance of evaluating reasoning trajectories according to their contribution to downstream recommendation performance rather than treating all trajectories equally. 
Moreover, removing either the \gen (\textit{w/o \gen}) or the \rec objective (\textit{w/o \rec}) consistently degrades recommendation accuracy, indicating that jointly learning to distinguish both beneficial and detrimental reasoning is important for improving the reasoning generator and the recommendation encoder.

\begin{table*}[htbp]
  \centering
  \caption{Ablation study on key components (\textbf{Backbone}: {Qwen2.5-3B}; \textbf{Datasets}: {CDs and Vinyl}).}
  \vskip -0.1in
  \scalebox{0.7}{
    \begin{tabular}{|l|cccccc|}
    \toprule
    \textbf{Datasets} & \multicolumn{6}{c|}{\textbf{CDs and Vinyl}} \\
    \midrule
    \textbf{Methods} & \multicolumn{1}{c|}{\textbf{N@5}} & \multicolumn{1}{c|}{\textbf{N@10}} & \multicolumn{1}{c|}{\textbf{N@20}} & \multicolumn{1}{c|}{\textbf{H@5}} & \multicolumn{1}{c|}{\textbf{H@10}} & \textbf{H@20} \\
    \midrule
    \midrule
    GPT   & 0.0265±0.0004 & 0.0323±0.0007 & 0.0376±0.0002 & 0.0372±0.0014 & 0.0552±0.0011 & 0.0764±0.0014 \\
    Gemini & 0.0243±0.0011 & 0.0299±0.0010 & 0.0356±0.0012 & 0.0333±0.0010 & 0.0510±0.0009 & 0.0743±0.0013 \\
    DeepSeek & 0.0258±0.0031 & 0.0316±0.0036 & 0.0372±0.0031 & 0.0388±0.0048 & 0.0563±0.0059 & 0.0783±0.0053 \\
    \midrule
    \textit{w/o SE} & 0.0258±0.0021 & 0.0305±0.0032 & 0.0380±0.0020 & 0.0363±0.0035 & 0.0510±0.0069 & 0.0803±0.0013 \\
    \textit{Rand}  & 0.0269±0.0020 & 0.0327±0.0020 & 0.0373±0.0021 & 0.0384±0.0023 & 0.0559±0.0020 & 0.0743±0.0025 \\
    \textit{w/o \gen} & 0.0257±0.0031 & 0.0297±0.0039 & 0.0355±0.0035 & 0.0356±0.0045 & 0.0482±0.0069 & 0.0710±0.0060 \\
    \textit{w/o \rec} & 0.0262±0.0010 & 0.0322±0.0009 & 0.0395±0.0009 & 0.0381±0.0018 & 0.0568±0.0025 & \textbf{0.0825±0.0042} \\
    \textbf{\method} & \textbf{0.0295±0.0009} & \textbf{0.0348±0.0017} & \textbf{0.0404±0.0010} & \textbf{0.0421±0.0019} & \textbf{0.0584±0.0041} & 0.0804±0.0018 \\
    \bottomrule
    \end{tabular}%
    }
  \label{tab:ablation_study_cds}%
\end{table*}%

\begin{table}[htbp]
  \centering
  \caption{Ablation study on key components (\textbf{Backbone}: {Qwen2.5-3B}; \textbf{Datasets}: {Musical Instruments}).}
  \scalebox{0.7}{
    \begin{tabular}{|l|cccccc|}
    \toprule
    \textbf{Datasets} & \multicolumn{6}{c|}{\textbf{Musical Instruments}} \\
    \midrule
    \textbf{Methods} & \multicolumn{1}{c|}{\textbf{N@5}} & \multicolumn{1}{c|}{\textbf{N@10}} & \multicolumn{1}{c|}{\textbf{N@20}} & \multicolumn{1}{c|}{\textbf{H@5}} & \multicolumn{1}{c|}{\textbf{H@10}} & \textbf{H@20} \\
    \midrule
    \midrule
    GPT   & 0.0155±0.0004 & 0.0200±0.0002 & 0.0264±0.0002 & 0.0250±0.0011 & 0.0388±0.0009 & 0.0645±0.0006 \\
    Gemini & 0.0141±0.0004 & 0.0193±0.0012 & 0.0236±0.0014 & 0.0213±0.0006 & 0.0377±0.0032 & 0.0549±0.0039 \\
    DeepSeek & 0.0146±0.0020 & 0.0188±0.0033 & 0.0238±0.0036 & 0.0221±0.0044 & 0.0353±0.0086 & 0.0554±0.0100 \\
    \midrule
    \textit{w/o SE} & \textbf{0.0195±0.0006} & 0.0231±0.0013 & 0.0318±0.0012 & 0.0282±0.0008 & 0.0456±0.0011 & 0.0719±0.0008 \\
    \textit{Rand}  & 0.0165±0.0006 & 0.0218±0.0008 & 0.0279±0.0008 & 0.0243±0.0008 & 0.0404±0.0031 & 0.0645±0.0019 \\
    \textit{w/o \gen} & 0.0168±0.0019 & 0.0217±0.0023 & 0.0285±0.0020 & 0.0262±0.0030 & 0.0414±0.0039 & 0.0684±0.0029 \\
    \textit{w/o \rec} & 0.0139±0.0006 & 0.0199±0.0014 & 0.0258±0.0008 & 0.0214±0.0009 & 0.0400±0.0031 & 0.0639±0.0008 \\
    \textbf{\method} & 0.0194±0.0008 & \textbf{0.0255±0.0005} & \textbf{0.0323±0.0013} & \textbf{0.0290±0.0011} & \textbf{0.0480±0.0008} & \textbf{0.0755±0.0030} \\
    \bottomrule
    \end{tabular}%
    }
  \label{tab:ablation_study_music}%
\end{table}%

\subsection{Model Analysis}

\noindent
\textbf{\textit{4.4.1 Parameter Analysis.}}
\begin{figure}[t]
    \centering
    \includegraphics[width=0.8\linewidth,height=1.65in]{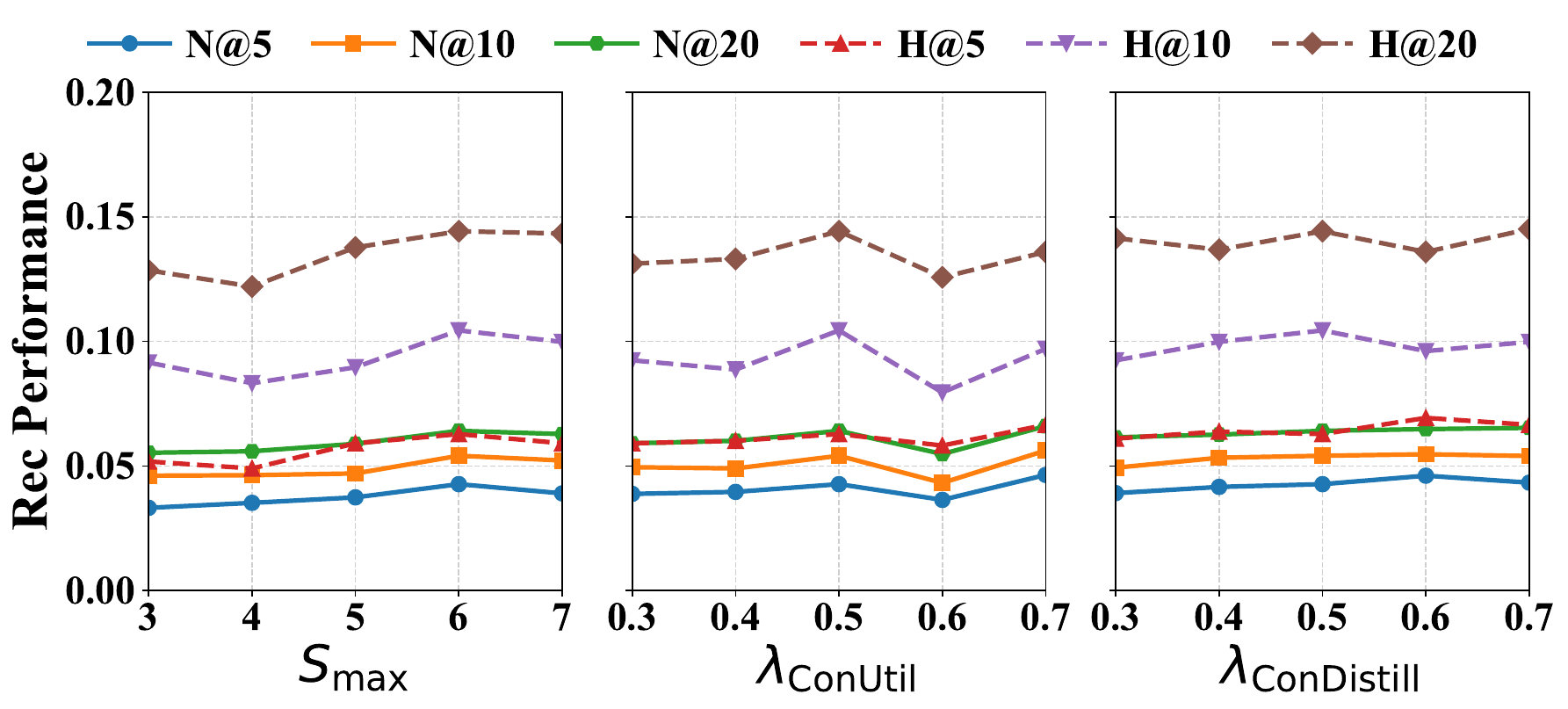}
    \caption{Effect of the hyper-parameters.}
    \label{fig:parameter}
    \vskip -0.2in
\end{figure}
To investigate the sensitivity of \method to different hyperparameters, we vary one hyperparameter at a time while keeping the others fixed and evaluate the resulting recommendation performance. The experiments are conducted on the Video Games dataset with Qwen2.5-3B as the backbone model.
As shown in Figure~\ref{fig:parameter}, increasing the maximum number of student reasoning candidates $S_{\max}$ from a small value generally improves recommendation performance, with strong results observed around $S_{\max}=6$ and $S_{\max}=7$. 
A small $S_{\max}$ restricts self-exploration, preventing the student from sufficiently exploring potentially more recommendation-beneficial reasoning trajectories and thereby leaving it constrained by the knowledge ceiling of the teacher models. 
In practice, $S_{\max}$ should therefore be selected according to the teacher's capability to balance teacher-guided distillation and student self-exploration. 
Stronger teachers require a smaller $S_{\max}$ for sufficient supervision, whereas weaker teachers benefit from a larger $S_{\max}$ to promote self-exploration and reduce teacher dependence.
For $\lambda_{\text{\rec}}$ and $\lambda_{\text{\gen}}$, the recommendation performance exhibits only moderate fluctuations within the tested ranges, without substantial degradation under different settings. 
This demonstrates that \method is robust to the weighting of the two contrastive objectives and does not require excessively fine-grained hyperparameter tuning to achieve competitive performance.

\begin{figure}[htbp]
    \vskip -0.15in
    \centering
    \includegraphics[width=0.8\linewidth,height=1.6in]{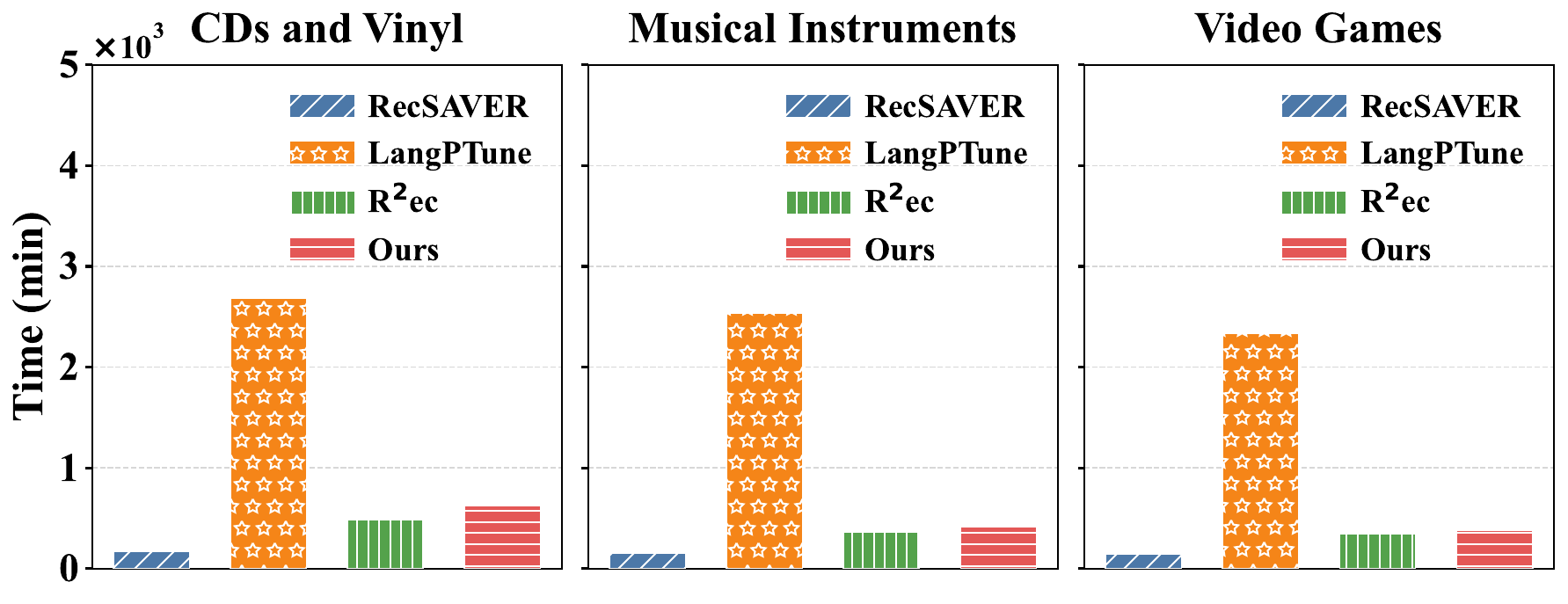  }
    \vskip -0.15in
    \caption{Training time of various methods.}
    \label{fig:training_time}
    \vskip -0.15in
\end{figure}

\noindent
\textbf{\textit{4.4.2 Training and Inference Efficiency.}}
\noindent
\textit{(1) Training Efficiency.}
Figure~\ref{fig:training_time} demonstrates that \method does not introduce prohibitive training overhead despite employing multiple teacher LLMs and student self-exploration. 
The reason is that all teacher-generated reasoning trajectories are produced in advance and cached for reuse, avoiding repeated teacher inference during iterative training. 
Meanwhile, Rank-Gain Utility Assessment only uses the lightweight recommendation encoder to perform candidate ranking, without introducing an additional LLM-based evaluator. 
Consequently, \method is substantially more efficient than LangPTune and incurs comparable computational overhead to R$^2$ec, while consistently achieving superior recommendation performance.

\begin{figure}[htbp]
    \vskip -0.15in
    \centering
    \includegraphics[width=0.8\linewidth,height=1.6in]{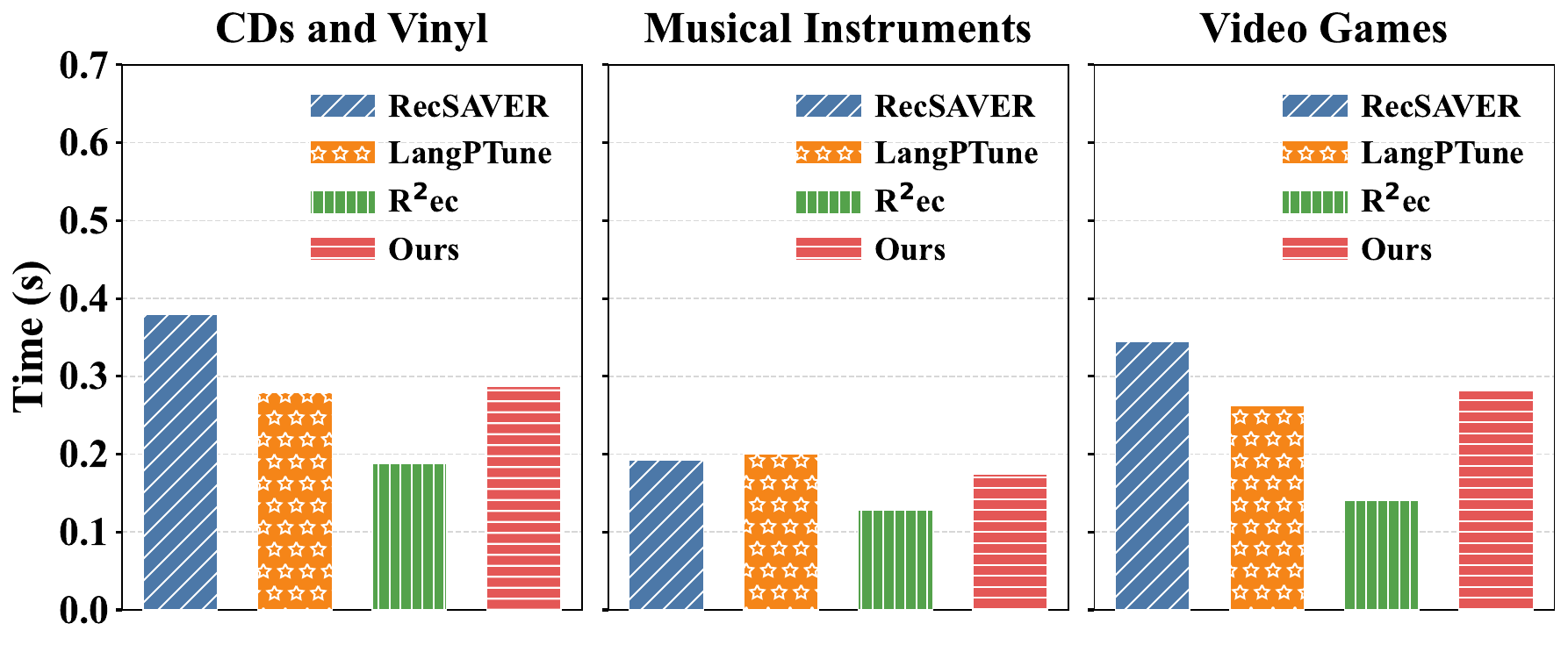}
    \vskip -0.05in
    \caption{Recommendation time of various methods.}
    \label{fig:rec_time}
    \vskip -0.15in
\end{figure}

\noindent
\textit{(2) Inference Efficiency.}
We further compare the inference efficiency of different reasoning-augmented recommender systems. 
All methods are deployed with the same vLLM inference engine~\cite{kwon2023efficient} and evaluated on identical hardware. 
As shown in Figure~\ref{fig:rec_time}, our method achieves competitive inference latency across all three datasets. 
While R$^2$ec exhibits the lowest inference latency, our method consistently delivers substantially better recommendation performance, as demonstrated in Table~\ref{tab:main_result_cds}, Table~\ref{tab:main_result_music}, and Table~\ref{tab:main_result_video} (\textbf{Appendix~\ref{appendix:main_experiments}}). 
These results indicate that the proposed framework introduces little additional inference overhead, while providing a favorable trade-off between recommendation accuracy and inference efficiency.

\begin{figure}[htbp]
    \vskip -0.15in
    \centering
    \includegraphics[width=0.8\linewidth]{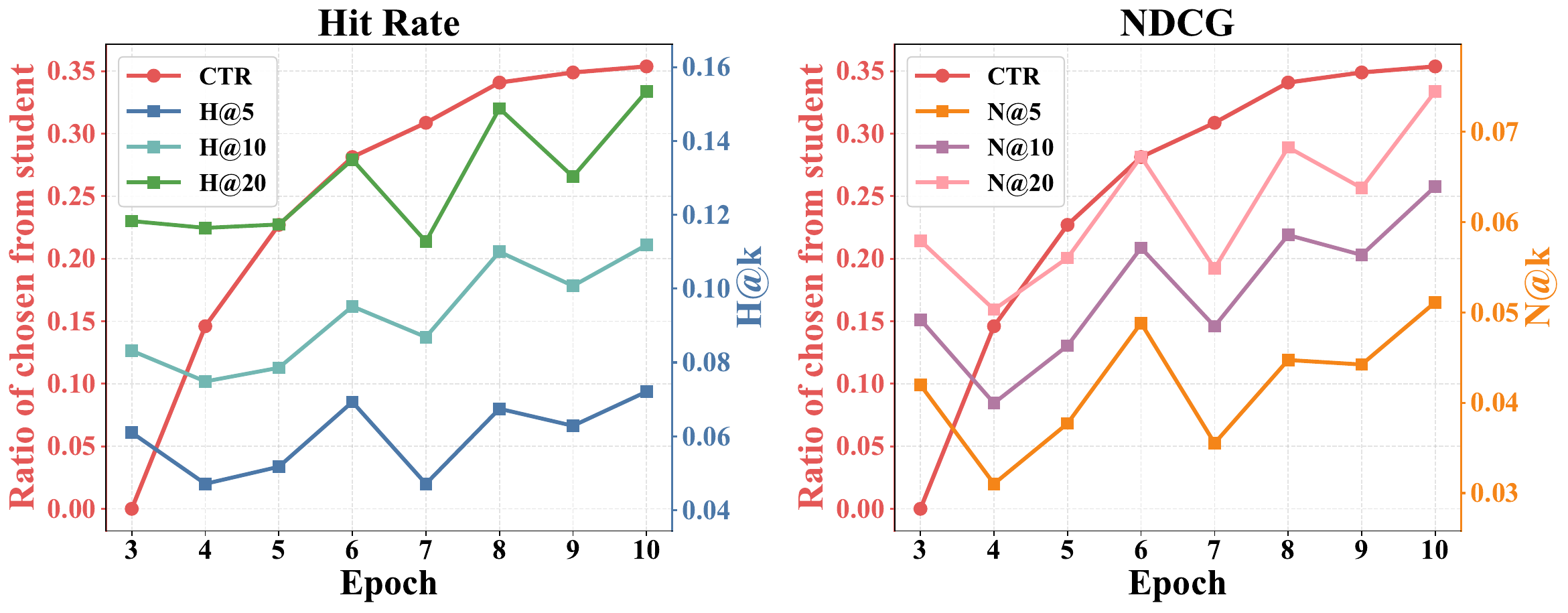}
    \vskip -0.1in
    \caption{Validation performance and student reasoning selection ratio during training.}
    \label{fig:traing_log}
\end{figure}

\noindent
\textbf{\textit{4.4.3 Effectiveness of Progressive Self-Exploration.}}
Figure~\ref{fig:traing_log} illustrates the evolution of the proposed self-exploration strategy during training. 
The experiment is conducted on the Video Games dataset using Qwen2.5-3B as the student model.
The red curve (CTR) reports the ratio of reasoning trajectories selected from the student model.
At the beginning of training, only a small proportion of student-generated reasoning is selected because the student has not yet acquired sufficient recommendation-specific reasoning capability. 
As training progresses, the student continuously improves by learning from recommendation-beneficial reasoning distilled from multiple teachers, enabling it to generate increasingly informative reasoning trajectories. 
Consequently, \select gradually selects more student-generated reasoning instead of teacher-generated reasoning for supervision. 
Meanwhile, the validation performance consistently improves across all metrics, indicating that the progressively explored reasoning provides increasingly effective guidance for downstream recommendation.

\noindent
To further verify that the effectiveness of \method is not merely attributable to using more reasoning trajectories, we compare it with MTMR, where each teacher generates multiple trajectories without student self-exploration, while keeping the total number of candidate trajectories identical to that of \method. 
As shown in Table~\ref{tab:multi_teacher_mutli_reasoning}, \method outperforms MTMR on most metrics, with particularly clear improvements on CDs and Vinyl and Video Games. 
These results indicate that increasing the quantity of teacher-generated reasoning alone is insufficient. Instead, student self-exploration introduces complementary reasoning beyond teacher imitation, thereby helping the student overcome the teacher knowledge ceiling.

\begin{table}[htbp]
  \centering
  \vskip -0.15in
  \caption{The overall performance of MTMR and \method.}
  \setlength{\tabcolsep}{1pt}
  \scalebox{0.7}{
    \begin{tabular}{|c|c|c|c|c|c|c|}
    \toprule
    \textbf{Methods} & \multicolumn{1}{c|}{\textbf{N@5}} & \multicolumn{1}{c|}{\textbf{N@10}} & \multicolumn{1}{c|}{\textbf{N@20}} & \multicolumn{1}{c|}{\textbf{H@5}} & \multicolumn{1}{c|}{\textbf{H@10}} & \textbf{H@20} \\
    \midrule
    \rowcolor[rgb]{ .929,  .929,  .929} \multicolumn{7}{|c|}{\textbf{CDs and Vinyl}} \\
    \midrule
    MTMR  & 0.0262±0.0010 & 0.0327±0.0026 & 0.0394±0.0033 & 0.0367±0.0004 & 0.0566±0.0049 & \textbf{0.0832±0.0079} \\
    \textbf{\method} & \textbf{0.0295±0.0009} & \textbf{0.0348±0.0017} & \textbf{0.0404±0.0010} & \textbf{0.0421±0.0019} & \textbf{0.0584±0.0041} & 0.0804±0.0018 \\
    \midrule
    \rowcolor[rgb]{ .929,  .929,  .929} \multicolumn{7}{|c|}{\textbf{Musical Instruments}} \\
    \midrule
    MTMR  & 0.0192±0.0010 & 0.0253±0.0012 & 0.0320±0.0019 & \textbf{0.0295±0.0020} & 0.0473±0.0028 & 0.0735±0.0060 \\
    \textbf{\method} & \textbf{0.0194±0.0008} & \textbf{0.0255±0.0005} & \textbf{0.0323±0.0013} & 0.0290±0.0011 & \textbf{0.0480±0.0008} & \textbf{0.0755±0.0030} \\
    \midrule
    \rowcolor[rgb]{ .929,  .929,  .929} \multicolumn{7}{|c|}{\textbf{Video Games}} \\
    \midrule
    MTMR  & \textbf{0.0438±0.0032} & 0.0525±0.0023 & 0.0626±0.0025 & \textbf{0.0638±0.0056} & 0.0912±0.0028 & 0.1313±0.0058 \\
    \textbf{\method} & 0.0421±0.0005 & \textbf{0.0536±0.0006} & \textbf{0.0632±0.0009} & 0.0626±0.0006 & \textbf{0.1020±0.0023} & \textbf{0.1399±0.0040} \\
    \bottomrule
    \end{tabular}%
    }
    \vskip -0.2in
  \label{tab:multi_teacher_mutli_reasoning}%
\end{table}%

\section{Related Work}
This section reviews recent advances in reasoning-augmented recommendation. 
According to their training paradigms, existing approaches can be broadly divided into three categories: SFT distillation-based methods, reinforcement learning (RL)-based methods, and hybrid approaches that combine both paradigms. 
Due to space limitations, the detailed and comprehensive review of the related work is provided in \textbf{Appendix~\ref{appendix:related_work}}.

\section{Conclusion}
In this paper, we presented \textbf{\method}, an adaptive reasoning distillation framework for reasoning-augmented recommendation. 
Unlike existing methods that directly imitate reasoning generated by a single fixed teacher, 
\method introduces multiple teachers and progressively expands the reasoning space through self-exploration, enabling the student to draw on the complementary strengths of different teachers and gradually overcome the knowledge limitation of general-purpose teachers. 
Furthermore, the proposed \selectfull (\select) adaptively identifies recommendation-beneficial reasoning, and our \contrastive objectives can jointly improve both reasoning generation and reasoning utilization by explicitly contrasting beneficial and detrimental reasoning trajectories for high-quality reasoning supervision. 
Extensive experiments on three real-world recommendation datasets demonstrate that \method consistently achieves superior recommendation performance, revealing the significant yet largely untapped potential of reasoning distillation for recommender systems.

\bibliographystyle{ieeenat_fullname}
\bibliography{Sections/Reference}

\newpage
\appendix
\section{Whole Process of \method}

\begin{algorithm}[htbp]
\caption{Training and Inference Procedure of \method}
\label{al:framework}
\KwIn{
Training dataset $\mathcal{D}$; teacher LLMs $\{\mathcal{T}_m\}_{m=1}^{M}$;\\
student reasoning generator $\mathcal{G}_{\theta}$; recommendation encoder $\mathcal{E}_{\phi}$.
}
\KwOut{Trained reasoning generator $\mathcal{G}_{\theta}$ and recommendation encoder $\mathcal{E}_{\phi}$.}

\textbf{\textcolor{gray}{// Training Phase}}\;

Generate teacher reasoning trajectories
$\mathcal{C}^{\mathrm{tea}}$
using all teacher LLMs\;

\For{epoch $=1,\ldots,E$}{

Generate $S_e$ self-exploration reasoning trajectories using the current student according to Eq.~\eqref{eq:self_schedule}\;

Construct the candidate reasoning pool
$\mathcal{C}^{(e)}
=
\mathcal{C}^{\mathrm{tea}}
\cup
\mathcal{C}^{\mathrm{stu}}$\;

Compute the \score for all candidates and obtain
$r^{+}$ and $r^{-}$ according to Eq.~\eqref{eq:best_worst}\;

Update the reasoning generator $\mathcal{G}_{\theta}$ by minimizing
$\mathcal{L}_{\mathrm{gen}}$
(Eq.~\eqref{eq:gen_total})\;

Update the recommendation encoder $\mathcal{E}_{\phi}$ by minimizing
$\mathcal{L}_{\mathrm{rec}}$
(Eq.~\eqref{eq:rec_total})\;


}

\vspace{0.05cm}
\textbf{\textcolor{gray}{// Inference Phase}}\;

Generate the user preference reasoning
$\hat r=\mathcal{G}_{\theta}(x)$\;

Encode the user representation
$\hat{\mathbf z}=\mathcal{E}_{\phi}([x;\hat r])$\;

Rank all candidate items according to Eq.~\eqref{eq:final_rank}
and return the Top-$K$ recommendation list\;

\end{algorithm}
\section{Experiments}

\subsection{Baselines}\label{appendix:baselines}
\begin{itemize}[leftmargin=*]
    \item \textbf{Caser}~\cite{tang2018personalized} models users' sequential behaviors using convolutional neural networks (CNNs), capturing high-order sequential patterns for next-item recommendation. 
    \item \textbf{GRU4Rec}~\cite{hidasi2015session} employs gated recurrent units (GRUs) to model users' sequential interaction histories, making it one of the earliest RNN-based sequential recommendation methods.
    \item \textbf{SASRec}~\cite{kang2018self} adopts the Transformer self-attention mechanism to capture long-range dependencies within user interaction sequences for personalized recommendation.
    \item \textbf{TIGER}~\cite{rajput2023recommender}	reformulates recommendation as a generative retrieval task, directly generating item identifiers using a T5-based sequence-to-sequence framework.
    \item \textbf{BigRec}~\cite{bao2025bi} introduces a bi-step grounding paradigm that first grounds user preferences and then performs recommendation with LLMs to improve recommendation accuracy.
    \item \textbf{D$^3$}~\cite{bao2024decoding} mitigates amplification bias and recommendation homogeneity in LLM-based recommender systems through a decoding strategy that promotes recommendation diversity.
    \item \textbf{RecSAVER}~\cite{tsai2024leveraging} enhances recommendation by prompting LLMs to explicitly generate preference reasoning for knowledge distillation, improving interpretability and recommendation accuracy. We use DeepSeek-V3.2-Thinking as the fixed teacher to generate reasoning trajectories and GTE-large-en-v1.5 as the recommendation encoder for sequential recommendations.
    \item \textbf{LangPTune}~\cite{gao2024end} adopts Reinforcement Learning for System Optimization (RLSO), a representative reinforcement learning paradigm, to enhance the reasoning capability of LLM-based recommender systems.
    \item \textbf{R$^2$ec}~\cite{you2026r} incorporates the recently proposed Group Relative Policy Optimization (GRPO)~\cite{guo2025deepseek} algorithm into LLM-empowered recommender systems to enhance user preference reasoning through reinforcement learning. 
\end{itemize}

\subsection{Implementation}\label{appendix:implementation}
All experiments are conducted on 2 $\times$ NVIDIA H20 GPUs using the AdamW optimizer. We adopt component-specific learning rates of $3\times10^{-6}$ for the recommendation encoder and $1\times10^{-5}$ for the student reasoning generator. Training is performed for at most 50 epochs, with early stopping based on the validation performance with a patience of 6 epochs.
For multi-teacher distillation, we employ three general-purpose teacher LLMs, namely \textit{DeepSeek-V3.2-Thinking}, \textit{Gemini-3-Flash-Preview}, and \textit{GPT-5-Mini}, each of which generates one preference reasoning trajectory for every training user. All teacher trajectories are generated with a temperature of $0$ and cached for reuse throughout training. 
The reasoning generator is fine-tuned using LoRA adapters~\cite{ding2023parameter} applied to all attention projections, feed-forward projections, and the language modeling head, while all parameters of the recommendation encoder are fully optimized. 
During training, the recommendation encoder is optimized using user representations constructed from the current student-generated reasoning trajectories, thereby avoiding the train-test discrepancy that would otherwise arise from using teacher-generated reasoning during training but student-generated reasoning at inference time. 
The two components are trained in an alternating manner until convergence.
The first two epochs serve as a warm-up stage, during which only the student reasoning generator is optimized via supervised fine-tuning on the best teacher trajectories, while the recommendation encoder remains frozen. 
After the warm-up stage, self-exploration is progressively activated. Specifically, we set $S_{\max}=6$ and $E_{\mathrm{s}}=10$, and increase the number of self-exploration trajectories $S_e$ linearly from $0$ to $S_{\max}$ as training proceeds.
For recommendation learning, we adopt a temperature of $\tau_r=0.1$ and set the weight of the \rec to $\lambda_{\mathrm{\rec}}=0.5$. For reasoning distillation, the weight of the \gen objective is set to $\lambda_{\mathrm{\gen}}=0.5$. 
We use a global batch size of 128, evenly partitioned across two GPUs for data-parallel training. 
All experiments are repeated three times, and the mean and standard deviation are reported to ensure the robustness and reliability of the results.

\subsection{Additional Experiments}\label{appendix:main_experiments}
\noindent
Table~\ref{tab:main_result_video} presents additional experimental results on the Video Games dataset. 
\method consistently achieves the best performance across all six metrics with both Qwen2.5-3B and Gemma-2-2B as the backbone. 
In particular, compared with the strongest baseline R$^2$ec, \method improves H@10 from 0.0921 to 0.1020 under Qwen2.5-3B, corresponding to a relative improvement of 10.7\%. 
It also improves H@20 from 0.1254 to 0.1313 under Gemma-2-2B. 
These results further confirm that the effectiveness of \method generalizes consistently across different datasets and student backbones.

\begin{table}[htbp]
  \centering
  \caption{The overall performance of baselines and \method on the \textbf{Video Games} dataset.}
  \scalebox{0.7}{
    \begin{tabular}{|c|c|c|c|c|c|c|}
    \toprule
    \textbf{Datasets} & \multicolumn{6}{c|}{\textbf{Video Games}} \\
    \midrule
    \textbf{Methods} & \textbf{N@5} & \textbf{N@10} & \textbf{N@20} & \textbf{H@5} & \textbf{H@10} & \textbf{H@20} \\
    \midrule
    Caser & 0.0233±0.0006 & 0.0289±0.0002 & 0.0366±0.0003 & 0.0468±0.0014 & 0.0644±0.0010 & 0.0952±0.0009 \\
    GRU4Rec & 0.0083±0.0028 & 0.0117±0.0041 & 0.0145±0.0048 & 0.0139±0.0055 & 0.0243±0.0092 & 0.0356±0.0121 \\
    SASRec & 0.0171±0.0010 & 0.0229±0.0010 & 0.0289±0.0027 & 0.0280±0.0023 & 0.0462±0.0028 & 0.0702±0.0111 \\
    TIGER & 0.0226±0.0007 & 0.0265±0.0005 & 0.0315±0.0007 & 0.0328±0.0027 & 0.0433±0.0016 & 0.0662±0.0031 \\
    \midrule
    \rowcolor[rgb]{ .929,  .929,  .929} \multicolumn{7}{|c|}{\textbf{Qwen2.5-3B}} \\
    \midrule
    D$^3$    & 0.0151±0.0006 & 0.0225±0.0013 & 0.0311±0.0059 & 0.0256±0.0023 & 0.0487±0.0067 & 0.0832±0.0246 \\
    BigRec & 0.0025±0.0010 & 0.0035±0.0005 & 0.0053±0.0005 & 0.0032±0.0005 & 0.0084±0.0010 & 0.0195±0.0050 \\
    RecSAVER & 0.0305±0.0033 & 0.0382±0.0030 & 0.0457±0.0044 & 0.0465±0.0030 & 0.0705±0.0035 & 0.1005±0.0077 \\
    LangPTune & 0.0271±0.0013 & 0.0360±0.0007 & 0.0448±0.0010 & 0.0422±0.0028 & 0.0699±0.0023 & 0.1050±0.0028 \\
    R$^2$ec  & \cellcolor[rgb]{ .851,  .882,  .949}\underline{0.0419±0.0002} & \cellcolor[rgb]{ .851,  .882,  .949}\underline{0.0515±0.0017} & \cellcolor[rgb]{ .851,  .882,  .949}\underline{0.0621±0.0013} & \cellcolor[rgb]{ .851,  .882,  .949}\underline{0.0622±0.0023} & \cellcolor[rgb]{ .851,  .882,  .949}\underline{0.0921±0.0023} & \cellcolor[rgb]{ .851,  .882,  .949}\underline{0.1346±0.0019} \\
    \textbf{\method}  & \cellcolor[rgb]{ .965,  .792,  .745}\textbf{0.0421±0.0005} & \cellcolor[rgb]{ .965,  .792,  .745}\textbf{0.0536±0.0006} & \cellcolor[rgb]{ .965,  .792,  .745}\textbf{0.0632±0.0009} & \cellcolor[rgb]{ .965,  .792,  .745}\textbf{0.0626±0.0006} & \cellcolor[rgb]{ .965,  .792,  .745}\textbf{0.1020±0.0023} & \cellcolor[rgb]{ .965,  .792,  .745}\textbf{0.1399±0.0040} \\
    \midrule
    \rowcolor[rgb]{ .929,  .929,  .929} \multicolumn{7}{|c|}{\textbf{Gemma-2-2B}} \\
    \midrule
    D$^3$    & 0.0286±0.0034 & 0.0383±0.0034 & 0.0513±0.0021 & 0.0416±0.0018 & 0.0718±0.0019 & 0.1229±0.0042 \\
    BigRec & 0.0131±0.0020 & 0.0155±0.0019 & 0.0189±0.0026 & 0.0160±0.0011 & 0.0241±0.0021 & 0.0354±0.0019 \\
    RecSAVER & 0.0293±0.0031 & 0.0358±0.0031 & 0.0433±0.0019 & 0.0431±0.0014 & 0.0631±0.0014 & 0.0933±0.0076 \\
    LangPTune & 0.0266±0.0020 & 0.0338±0.0021 & 0.0425±0.0016 & 0.0425±0.0009 & 0.0647±0.0019 & 0.0995±0.0028 \\
    R$^2$ec  & \cellcolor[rgb]{ .851,  .882,  .949}\underline{0.0380±0.0016} & \cellcolor[rgb]{ .851,  .882,  .949}\underline{0.0483±0.0010} & \cellcolor[rgb]{ .851,  .882,  .949}\underline{0.0574±0.0015} & \cellcolor[rgb]{ .851,  .882,  .949}\underline{0.0557±0.0022} & \cellcolor[rgb]{ .851,  .882,  .949}\underline{0.0890±0.0023} & \cellcolor[rgb]{ .851,  .882,  .949}\underline{0.1254±0.0043} \\
    \textbf{\method}  & \cellcolor[rgb]{ .965,  .792,  .745}\textbf{0.0387±0.0010} & \cellcolor[rgb]{ .965,  .792,  .745}\textbf{0.0492±0.0015} & \cellcolor[rgb]{ .965,  .792,  .745}\textbf{0.0595±0.0027} & \cellcolor[rgb]{ .965,  .792,  .745}\textbf{0.0573±0.0016} & \cellcolor[rgb]{ .965,  .792,  .745}\textbf{0.0902±0.0019} & \cellcolor[rgb]{ .965,  .792,  .745}\textbf{0.1313±0.0074} \\
    \bottomrule
    \end{tabular}%
    }
  \label{tab:main_result_video}%
\end{table}%

\subsection{Additional Ablation Study}\label{appendix:ablation_study}
\noindent
Table~\ref{tab:ablation_study_video} reports the additional ablation results on the Video Games dataset.
The complete \method consistently outperforms all single-teacher and component-removed variants across every evaluation metric. 
For example, replacing \select with random trajectory selection also decreases H@10 from 0.1020 to 0.0844, demonstrating the importance of selecting reasoning trajectories according to their downstream recommendation utility. 
Moreover, the performance degradation caused by removing self-exploration, \gen or \rec further confirms that all proposed components contribute positively to the final recommendation performance.

\begin{table}[htbp]
  \centering
  \vskip -0.1in
  \caption{Ablation study on key components (\textbf{Backbone}: {Qwen2.5-3B}; \textbf{Dataset}: {Video Games}).}
  \scalebox{0.7}{
    \begin{tabular}{|l|c|c|c|c|c|c|}
    \toprule
    \textbf{Datasets} & \multicolumn{6}{c|}{\textbf{Video Games}} \\
    \midrule
    \textbf{Methods} & \multicolumn{1}{c|}{\textbf{N@5}} & \multicolumn{1}{c|}{\textbf{N@10}} & \multicolumn{1}{c|}{\textbf{N@20}} & \multicolumn{1}{c|}{\textbf{H@5}} & \multicolumn{1}{c|}{\textbf{H@10}} & \textbf{H@20} \\
    \midrule
    \midrule
    GPT   & 0.0395±0.0005 & 0.0482±0.0007 & 0.0589±0.0019 & 0.0598±0.0021 & 0.0866±0.0010 & 0.1294±0.0097 \\
    Gemini & 0.0400±0.0008 & 0.0491±0.0029 & 0.0564±0.0043 & 0.0548±0.0021 & 0.0832±0.0066 & 0.1137±0.0145 \\
    DeepSeek & 0.0309±0.0053 & 0.0390±0.0056 & 0.0493±0.0057 & 0.0481±0.0065 & 0.0736±0.0076 & 0.1146±0.0080 \\
    \midrule
    \textit{w/o SE} & 0.0413±0.0006 & 0.0527±0.0006 & 0.0623±0.0007 & 0.0609±0.0008 & 0.0998±0.0021 & 0.1343±0.0030 \\
    \textit{Rand}  & 0.0376±0.0013 & 0.0457±0.0016 & 0.0567±0.0013 & 0.0591±0.0024 & 0.0844±0.0038 & 0.1284±0.0024 \\
    \textit{w/o \gen} & 0.0332±0.0007 & 0.0434±0.0016 & 0.0521±0.0015 & 0.0530±0.0019 & 0.0847±0.0033 & 0.1195±0.0044 \\
    \textit{w/o \rec} & 0.0390±0.0020 & 0.0496±0.0024 & 0.0602±0.0030 & 0.0596±0.0020 & 0.0915±0.0009 & 0.1334±0.0028 \\
    \textit{\method} & \textbf{0.0421±0.0005} & \textbf{0.0536±0.0006} & \textbf{0.0632±0.0009} & \textbf{0.0626±0.0006} & \textbf{0.1020±0.0023} & \textbf{0.1399±0.0040} \\
    \bottomrule
    \end{tabular}%
    }
  \label{tab:ablation_study_video}%
\end{table}%

\section{Used Prompts}\label{appendix:used_prompt}
The prompt used to guide the reasoning generator in producing user preference reasoning is shown in \textbf{Prompt~\ref{prompt:rec}}. 
Specifically, {\textbf{\{category\}}} denotes the product category (e.g., music, games, or musical instruments), while {\textbf{\{items\}}} contains the textual metadata of the items previously interacted with by the user, including the title, description, and categories. Here, description refers to a concise summary of each item. 
After the reasoning is generated, we append the generated reasoning to the prompt in the form of \textit{"User preference summary: {\textbf{\{Reasoning\}}}"}, and feed the resulting prompt into the recommendation encoder to generate the final recommendation.
\begin{figure}[htbp]
\mycommentbox{\textbf{Prompt 1: Instruction for the reasoning generator $\mathcal{G}_{\theta}$}}{prompt:rec}{

The user has rated (out of 5.0) the following {\textbf{\{category\}}} products arranged chronologically from the oldest (top) to the newest (bottom). 
Please provide a high-level summary of the user preference in detail. 
Here are the items the user interacted with and the rating from the user: {\textbf{\{items\}}}

}
\end{figure}

\section{Case Study}\label{appendix:reasoning_cases}
To qualitatively evaluate the effectiveness of \method, we present some representative examples in \textbf{Example~\ref{example:1} and \ref{example:2}}. 
As shown, \method consistently generates accurate and interpretable preference reasoning by identifying users' high-level interests from historical interactions, rather than merely summarizing individual items. 
For example, it successfully captures preferences for premium gaming peripherals (Example~1) and the PlayStation 2 ecosystem and classic open-world games (Example~2).
These reasoning trajectories effectively characterize the users' underlying preferences and provide informative guidance for the recommendation encoder, thereby facilitating correct recommendations. 
These qualitative examples further demonstrate that \method learns accurate reasoning that is both interpretable and beneficial for downstream recommendation.
\begin{figure*}[t]
\mycommentbox{\textbf{Example 1}}{example:1}{

\textbf{Recommendation Query: }

The user has rated the following Video Games products arranged chronologically from the oldest to the newest. Please provide a high-level summary of the user preference in detail. Here are the items the user interacted with and the rating from the user:

\vspace{0.5em}

\hspace{1em} --- 1. [EASARS Wireless Gaming Headset] (\colorbox{blue!10}{\small Rating: 4.0})

\hspace{1em} --- 2. [Smart Duck Keyboard] (\colorbox{blue!10}{\small Rating: 5.0})

\hspace{1em} --- 3. [AULA Mechanical Gaming Keyboard] (\colorbox{blue!10}{\small Rating: 5.0})

\hspace{1em} --- ...

\vspace{0.25em}

\textbf{Preference Reasoning: }

The user strongly prefers {\textbf{premium gaming peripherals}}, particularly high-quality {\textbf{mechanical keyboards}} with {\textbf{RGB lighting}}, {\textbf{hot-swappable switches}}, and strong {\textbf{typing performance}}. They also value immersive gaming audio, but are more tolerant of headsets than keyboards, indicating higher expectations for keyboard quality and customization.

}

\vskip 0.2in

\mycommentbox{\textbf{Example 2}}{example:2}{

\textbf{Recommendation Query: }

The user has rated the following Video Games products arranged chronologically from the oldest to the newest. Please provide a high-level summary of the user preference in detail. Here are the items the user interacted with and the rating from the user:

\vspace{0.5em}

\hspace{1em} --- 1. [Grand Theft Auto 3] 
(\colorbox{blue!10}{\small Rating: 5.0})

\hspace{1em} --- 2. [Grand Theft Auto Vice City] 
(\colorbox{blue!10}{\small Rating: 5.0})

\hspace{1em} --- 3. [PlayStation 2 Slim Console (Renewed)] 
(\colorbox{blue!10}{\small Rating: 4.0})

\hspace{1em} --- ...

\vspace{0.25em}

\textbf{Preference Reasoning: }

The user shows a strong preference for {\textbf{PlayStation 2 titles}}, especially {\textbf{classic open-world action games}}. They enjoy {\textbf{immersive single-player experiences}} and demonstrate {\textbf{nostalgia toward the PS2 ecosystem}}, while also maintaining interest in {\textbf{PS2 hardware}}.

}

\end{figure*}

\section{Related Work}\label{appendix:related_work}
\noindent
\textbf{SFT distillation-based methods.}
Motivated by the strong reasoning capabilities of LLMs, a growing body of research has explored transferring such capabilities to recommender systems through knowledge distillation for enhanced user preference modeling~\cite{wang2024can,tsai2024leveraging,wang2024rdrec,kim2025driven,qi2025knowledge,zhu2026learning,jiang2025scoter,dong2026reasoning,ao2026passing,shahgir2026scored,lu2026think}.
For example, 
SLIM~\cite{wang2024can} transfers the reasoning ability of a large LLM into a much smaller recommendation model through rationale distillation, so that the student can perform recommendations with explicit reasoning while remaining computationally efficient. 
Exp3rt~\cite{kim2025driven} trains the student to sequentially perform preference extraction, user/item profile construction, and step-by-step reasoning for rating prediction, enabling the model to leverage rich semantic information from reviews while producing interpretable recommendation rationales. 
The most closely related study is IRMD~\cite{lu2026think}, which synthesizes CoT trajectories and progressively distills them into the student’s hidden states to enable implicit reasoning without generating rationales at inference. 
Despite this connection, IRMD differs fundamentally from MUSE. 
It filters reasoning using teacher self-assessed correctness and coherence rather than its actual utility for downstream recommendation, and uses only teacher-generated reasoning as supervision without addressing the teacher knowledge ceiling. 
Moreover, its implicit distillation aligns the teacher’s and student’s final hidden states, requiring white-box access that is typically unavailable for commercial LLMs. 
Finally, although termed multi-teacher, its implementation uses only Qwen2.5-14B under three temperatures, whereas MUSE leverages heterogeneous teacher LLMs and progressive student self-exploration.

\noindent
\textbf{Reinforcement learning (RL)-based methods.}
With the remarkable success of reinforcement learning in enhancing the reasoning capabilities of LLMs across a wide range of complex tasks, an increasing number of studies have begun to leverage reinforcement learning algorithms to improve the reasoning ability of LLM-based recommender systems~\cite{huang2026rerec,you2026r,gao2024end,zhang2025reinforced,tang2026think,ding2026towards,jiang2026diffureason,pan2026flexrec,zheng2026sapo,gao2026reinforced}. 
For example, 
LangPTune~\cite{gao2024end} proposes an end-to-end framework that optimizes LLM-generated language-based user profiles for recommendation. Instead of relying on supervised reasoning distillation, it employs reinforcement learning to optimize the profile generator using downstream recommendation performance as the reward signal, while jointly training the ranking model through contrastive learning to improve recommendation accuracy and profile quality. 
R$^2$ec~\cite{you2026r} integrates reasoning generation and item prediction into a unified large recommender model, enabling end-to-end optimization of both tasks. It employs an RL-based optimization framework (RecPO) to jointly optimize reasoning and recommendation using a fused reward derived from recommendation supervision, enabling end-to-end training without human reasoning annotations.
ReRec~\cite{huang2026rerec} introduces fine-grained reward shaping together with reasoning-aware advantage estimation to provide more informative supervision for both recommendation outcomes and intermediate reasoning steps during reinforcement fine-tuning.

\noindent
\textbf{Hybrid methods.}
Beyond these approaches, several studies adopt hybrid training paradigms that integrate supervised fine-tuning and reinforcement learning to further improve the reasoning capabilities of LLM-empowered recommender systems~\cite{zhang2025slow,hu2026adnanny,liang2026gr2,chen2026lasar}.
RecOne~\cite{kong2026think} adopts a hybrid SFT+RL paradigm for reasoning-enhanced recommendation. It first initializes the LLM through supervised fine-tuning on a small set of high-quality reasoning samples, and subsequently improves its reasoning capabilities via reinforcement learning with recommendation-specific rewards.
GR2~\cite{liang2026gr2} further applies the hybrid training strategy to recommendation re-ranking, where the model is first supervised using high-quality hierarchical reasoning traces generated by a teacher LLM and then refined through a re-ranking-oriented reinforcement learning objective.

\end{document}